\documentclass[a4paper,11pt]{article}
\pdfoutput=1

\usepackage{jheppub}
\usepackage[numbers,sort&compress]{natbib}
\usepackage{graphicx}  
\usepackage{float}
\usepackage{dcolumn}   
\usepackage{bm}        
\usepackage{amssymb}   
\usepackage{slashed}   
\usepackage{amsmath}   
\usepackage{verbatim}  
\usepackage{color} 
\usepackage{hyperref}

\graphicspath{{Figures/}}
\begin{document}

\title{Tri-photon at muon collider: a new process to probe the anomalous quartic gauge couplings}

\author[a]{Ji-Chong Yang}
\emailAdd{yangjichong@lnnu.edu.cn}
\affiliation[a]{Department of Physics, Liaoning Normal University, Dalian 116029, China}

\author[a]{Zhi-Bin Qin}

\author[a]{Xue-Ying Han}

\author[a]{Yu-Chen Guo}
\emailAdd{ycguo@lnnu.edu.cn}

\author[b]{Tong Li}
\emailAdd{litong@nankai.edu.cn}
\affiliation[b]{School of Physics, Nankai University, Tianjin 300071, China}

\abstract{
The muon collider has recently received a great deal of attention because of its ability to achieve both high energy and high luminosity.
It plays as a gauge boson collider because the vector boson scattering (VBS) becomes the dominant production topology for Standard Model processes starting from a few TeV of collision energy.
In this paper, we propose that the process of $\mu^+\mu^-$ annihilation into tri-photon is also very sensitive to the search of anomalous quartic gauge couplings (aQGCs). We investigate the projected constraints on the transverse operators contributing to aQGCs through $\mu^+\mu^-\to Z^\ast/\gamma^\ast\to \gamma\gamma\gamma$ at muon colliders. For the muon collider with $\sqrt{s}=3$ TeV and $\mathcal{L}=1\;{\rm ab}^{-1}$, the expected constraints are about two orders of magnitude stronger than those at the 13 TeV LHC.
}

\maketitle

\section{\label{sec1}Introduction}

As the foundation of the Standard Model~(SM), the non-Abelian nature of $SU(2)\times U(1)$ gauge symmetry has received longstanding attention in both experiments and theoretical studies. The precision measurements of the self-couplings of electroweak (EW)
gauge bosons drive the test of the electroweak symmetry breaking mechanism and the search for new physics (NP) beyond the SM. Any deviation away from the SM gauge interactions and anomalous gauge couplings would indicate the existence of new physics above the EW scale in terms of higher dimensional effective operators.

In the SM effective field theory~(SMEFT)~\cite{weinberg,SMEFTReview1,SMEFTReview2,SMEFTReview3}, the gauge interactions can be extended by high-dimensional operators contributing to anomalous triple gauge couplings~(aTGCs) and anomalous quartic gauge couplings~(aQGCs)~\cite{aqgcold,aqgcnew}.
Although the SMEFT has been mainly applied to study dimension-6 operators, many groups recently pay much attention to the dimension-8 operators~\cite{d81,vbs1,ssww,wastudy,wwstudy,zastudy}.
The dimension-8 operators are important with respect to the convex geometry perspective to the operator space~\cite{positivity1,positivity2,positivity3}.
Moreover, there exist various NP models generating dimension-8 effective operators relevant for aQGCs~\cite{composite1,composite2,extradim,2hdm1,2hdm2,zprime1,zprime2,alp1,alp2,wprime}. The aQGCs can then be induced by tree-level diagrams while the aTGCs cannot~\cite{looportree}. In particular, there are distinct cases where dimension-6 operators are absent but the dimension-8 operators show up~\cite{bi1,bi2,bi3,ntgc1,ntgc2,ntgc3,ntgc4,ntgc5,ntgc6,ntgc7}.
The aQGCs due to the dimension-8 operators have been extensively studied at the Large Hadron Collider (LHC)~\cite{sswwexp1,sswwexp2,zaexp1,zaexp2,zaexp3,waexp1,zzexp1,zzexp2,wzexp1,wzexp2,wwexp1,wwexp2,wvzvexp,waexp2,zzexp3}.

Recently, the progress on the studies of muon collider has drawn a lot of attention~\cite{muoncollider1,muoncollider2,muoncollider4,muoncollider6,muoncollider7,muoncollider8,muoncollider3,muoncollider5,Han:2020uid,Han:2021kes,Aime:2022flm}.
Compared with the composite particles at hadron colliders, the muon collider provides a cleaner environment without suffering from QCD background. Also, the muon collider is able to reach both multi-TeV energies and a high integrated luminosity scaling with energy quadratically. As a result, the muon collider is undoubtedly an ideal place to search for new physics above the EW scale and precisely measure the aQGCs.
The aQGCs induced processes can be categorized into the vector boson scattering/fusion (VBS/VBF) processes and the tri-boson productions.
At the LHC, the VBS processes are generally more sensitive to the aQGCs than the tri-boson processes~\cite{triboson1,triboson2,triboson3,triboson4,triboson5}.
At the muon collider, the gauge bosons in final states of VBS processes are associated with very energetic muons or neutrinos in the forward region with respect to the beam. They become the dominant production mode for all SM processes starting from a few TeV energies due to the logarithmic enhancement from gauge boson radiation, which also holds for many new physics models~\cite{muoncollider3,Han:2020uid}.
On the other hand, the tri-boson processes are normally suppressed by $1/s$ in the s-channel propagator. In spite of this, it is expected that the tri-boson processes without forward objects carrying away energies can also be sensitive to the aQGCs induced by dimension-8 operators with four field strength tensors.
This is especially true for the tri-photon production
\begin{eqnarray}
\mu^+\mu^-\to Z^\ast/\gamma^\ast\to \gamma\gamma\gamma\;,
\end{eqnarray}
where there are no subsequent decays in final states. This channel was utilized to probe the axion-like particle at muon colliders through $\mu^+\mu^-\to Z^\ast/\gamma^\ast\to \gamma a (\to \gamma\gamma)$~\cite{Bao:2022onq,Han:2022mzp}. In this paper, we study the sensitivity of the tri-photon process to the neutral anomalous couplings $\gamma\gamma\gamma Z/\gamma$ and the relevant dimension-8 effective operators at muon colliders.

The rest of this paper is organized as follows.
In Sec.~\ref{sec2}, the dimension-8 operators contributing to aQGCs are introduced. We also discuss the tri-photon process and compare with the VBS.
The numerical results are presented in Sec.~\ref{sec3}.
In Sec.~\ref{sec4} we summarize our conclusions.

\section{\label{sec2}Anomalous quartic gauge couplings and tri-photon process}

\subsection{\label{sec2.1}A brief introduction of the anomalous quartic gauge couplings}

The dimension-8 operators frequently used in previous works can be classified as scalar/longitudinal operators $O_{S_i}$, mixed transverse and longitudinal operators $O_{M_i}$ and transverse operators $O_{T_i}$~\cite{vbscan}
\begin{eqnarray}
\mathcal{L}_{\rm aQGC}=\sum_{i=0}^{2}{f_{S_i}\over \Lambda^4}O_{S_i}+\sum_{i=0}^{7}{f_{M_i}\over \Lambda^4}O_{M_i}+\sum_{i=0}^{9}{f_{T_i}\over \Lambda^4}O_{T_i}\;,
\end{eqnarray}
where $f_{S_i}$, $f_{M_i}$ and $f_{T_i}$ are dimensionless coefficients, and $\Lambda$ is the cut-off scale.
$O_{S_i}$ and $O_{M_i}$ operators are irrelevant for the study of tri-photon channel in this paper. The relevant $O_{T_i}$ operators with four field strength tensors are~\cite{aqgcold,aqgcnew}
\begin{equation}
\begin{split}
&O_{T_0}={\rm Tr}\left[\widehat{W}_{\mu\nu}\widehat{W}^{\mu\nu}\right]\times {\rm Tr}\left[\widehat{W}_{\alpha\beta}\widehat{W}^{\alpha\beta}\right],\\
&O_{T_2}={\rm Tr}\left[\widehat{W}_{\alpha\mu}\widehat{W}^{\mu\beta}\right]\times {\rm Tr}\left[\widehat{W}_{\beta\nu}\widehat{W}^{\nu\alpha}\right],\\
&O_{T_6}={\rm Tr}\left[\widehat{W}_{\alpha\nu}\widehat{W}^{\mu\beta}\right]\times B_{\mu\beta}B^{\alpha\nu},\\
&O_{T_8}=B_{\mu\nu}B^{\mu\nu}\times B_{\alpha\beta}B^{\alpha\beta},\\
\end{split}
\quad
\begin{split}
&O_{T_1}={\rm Tr}\left[\widehat{W}_{\alpha\nu}\widehat{W}^{\mu\beta}\right]\times {\rm Tr}\left[\widehat{W}_{\mu\beta}\widehat{W}^{\alpha\nu}\right],\\
&O_{T_5}={\rm Tr}\left[\widehat{W}_{\mu\nu}\widehat{W}^{\mu\nu}\right]\times B_{\alpha\beta}B^{\alpha\beta},\\
&O_{T_7}={\rm Tr}\left[\widehat{W}_{\alpha\mu}\widehat{W}^{\mu\beta}\right]\times B_{\beta\nu}B^{\nu\alpha},\\
&O_{T_9}=B_{\alpha\mu}B^{\mu\beta}\times B_{\beta\nu}B^{\nu\alpha},\\
\end{split}
\label{eq.2.2}
\end{equation}
where $\widehat{W}\equiv \sum _i W^i \sigma ^i / 2$ with $\sigma^i$ being the Pauli matrices.
The constraints on the coefficients of these operators were obtained at the LHC, and listed in Table~\ref{tab.1}. These existing bounds are at the level of $f_{T_i}/\Lambda^4\simeq \mathcal{O}(0.1)$ TeV$^{-4}$.
Note that the real unitarity analysis was not performed when obtaining the constraints on coefficients in Table~\ref{tab.1}. Instead, Refs.~\cite{waexp2,zzexp3} provided the maximally allowed energy scales when the coefficients are equal to the constraints in Table~\ref{tab.1}. With unitarity taken into account, according to our previous studies~\cite{wastudy,zastudy,Yang:2020slb}, the LHC constraints should become weaker.

\begin{table}
\begin{center}
\begin{tabular}{c|c||c|c}
\hline
 coefficient & constraint & coefficient & constraint \\
\hline
 $f_{T_0}/\Lambda ^4\;({\rm TeV^{-4}})$ & $[-0.12, 0.11]$~\cite{wvzvexp} & $f_{T_6}/\Lambda ^4\;({\rm TeV^{-4}})$ & $[-0.4, 0.4]$~\cite{waexp2} \\
 $f_{T_1}/\Lambda ^4\;({\rm TeV^{-4}})$ & $[-0.12, 0.13]$~\cite{wvzvexp} & $f_{T_7}/\Lambda ^4\;({\rm TeV^{-4}})$ & $[-0.9, 0.9]$~\cite{waexp2} \\
 $f_{T_2}/\Lambda ^4\;({\rm TeV^{-4}})$ & $[-0.28, 0.28]$~\cite{wvzvexp} & $f_{T_8}/\Lambda ^4\;({\rm TeV^{-4}})$ & $[-0.43, 0.43]$~\cite{zzexp3} \\
 $f_{T_5}/\Lambda ^4\;({\rm TeV^{-4}})$ & $[-0.5, 0.5]$~\cite{waexp2} & $f_{T_9}/\Lambda ^4\;({\rm TeV^{-4}})$ & $[-0.92, 0.92]$~\cite{zzexp3} \\
\hline
\end{tabular}
\end{center}
\caption{The existing LHC constraints on the $O_{T_i}$ coefficients obtained at $95\%$ CL.}
\label{tab.1}
\end{table}

\subsection{\label{sec2.2}The annihilation process compared with the VBS processes}

At the muon collider, the annihilation and VBS processes used to study the aQGCs are
\begin{eqnarray}
&&\mu^+\mu^-\to V_1V_2V_3~~~({\rm annihilation})\;,\nonumber \\
&&\mu^+\mu^-\to f f' V_1V_2~~~({\rm VBS})\;,
\end{eqnarray}
where $f$ and $f'$ are the outgoing muons or neutrinos, and $V_i$ denotes SM gauge bosons.
An important difference between the muon collider and the $pp$ collider is that the annihilation process at $pp$ collider must be led by sea quark partons in initial states.
Nevertheless, before starting to study the tri-photon process, we need to compare the annihilation with the VBS process at the muon collider.
When a new particle $X$ is produced, there is a relative scaling~\cite{muoncollider5}
\begin{equation}
\begin{split}
&\frac{\sigma ^{\rm BSM}_{\rm VBF}}{\sigma ^{\rm BSM}_{\rm ann}}\propto \alpha _W^2 \frac{s}{m_X^2}\log^2\left(\frac{s}{m_V^2}\right)\log \left(\frac{s}{m_X^2}\right)\;,
\end{split}
\label{eq.2.3}
\end{equation}
where $\sigma ^{\rm BSM}_{\rm VBF}$ and $\sigma^{\rm BSM}_{\rm ann}$ are the cross-sections of VBF and annihilation processes with $X$ in the final states, $m_{V,X}$ are the masses of intermediate vector boson and the new particle $X$, respectively. Well above the threshold for $\sqrt{s}\gg m_V$, the initial muons substantially radiate massive EW gauge bosons under an approximately unbroken SM gauge symmetry. This leads to the above  double-logarithmic enhancement $\log^2\left(\frac{s}{m_V^2}\right)$ and thus the VBF process will dominate over the annihilation at high energies.

\begin{figure}
\begin{center}
\includegraphics[width=0.7\textwidth]{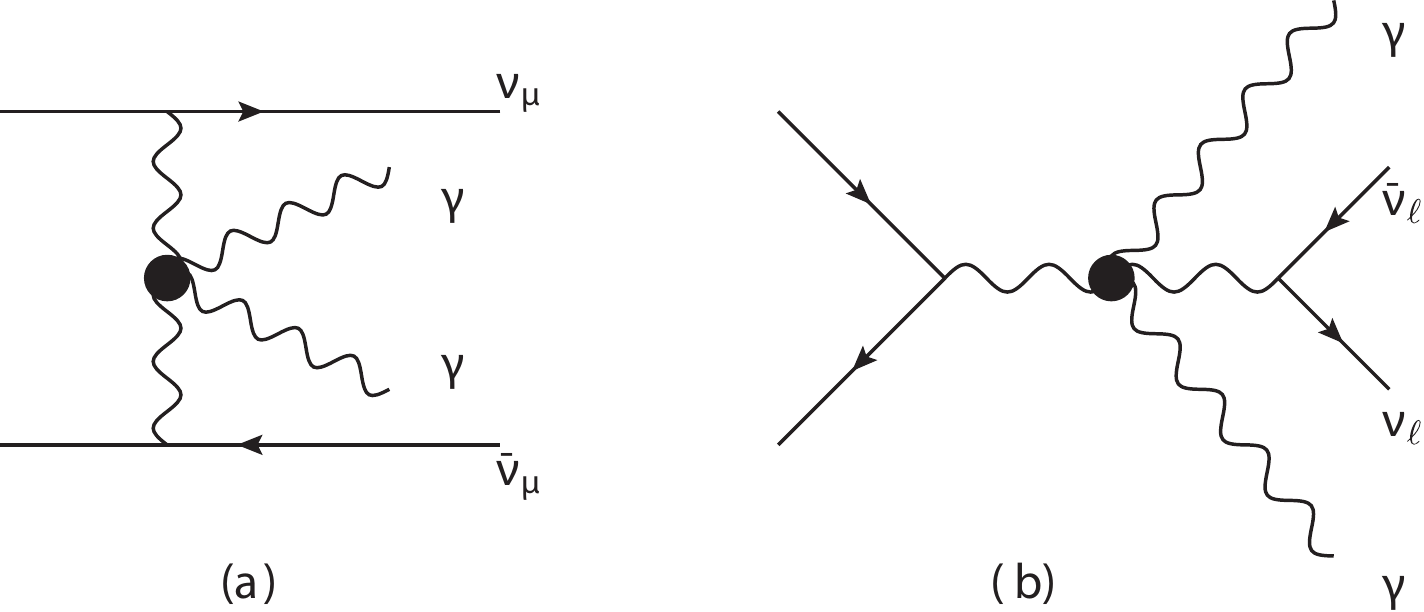}
\caption{\label{fig:llvvaadiagram}The typical Feynman diagrams of the aQGCs contribution to the process $\mu^+\mu^-\to \gamma\gamma \nu \bar{\nu}$.}
\end{center}
\end{figure}

\begin{figure}
\begin{center}
\includegraphics[width=0.48\textwidth]{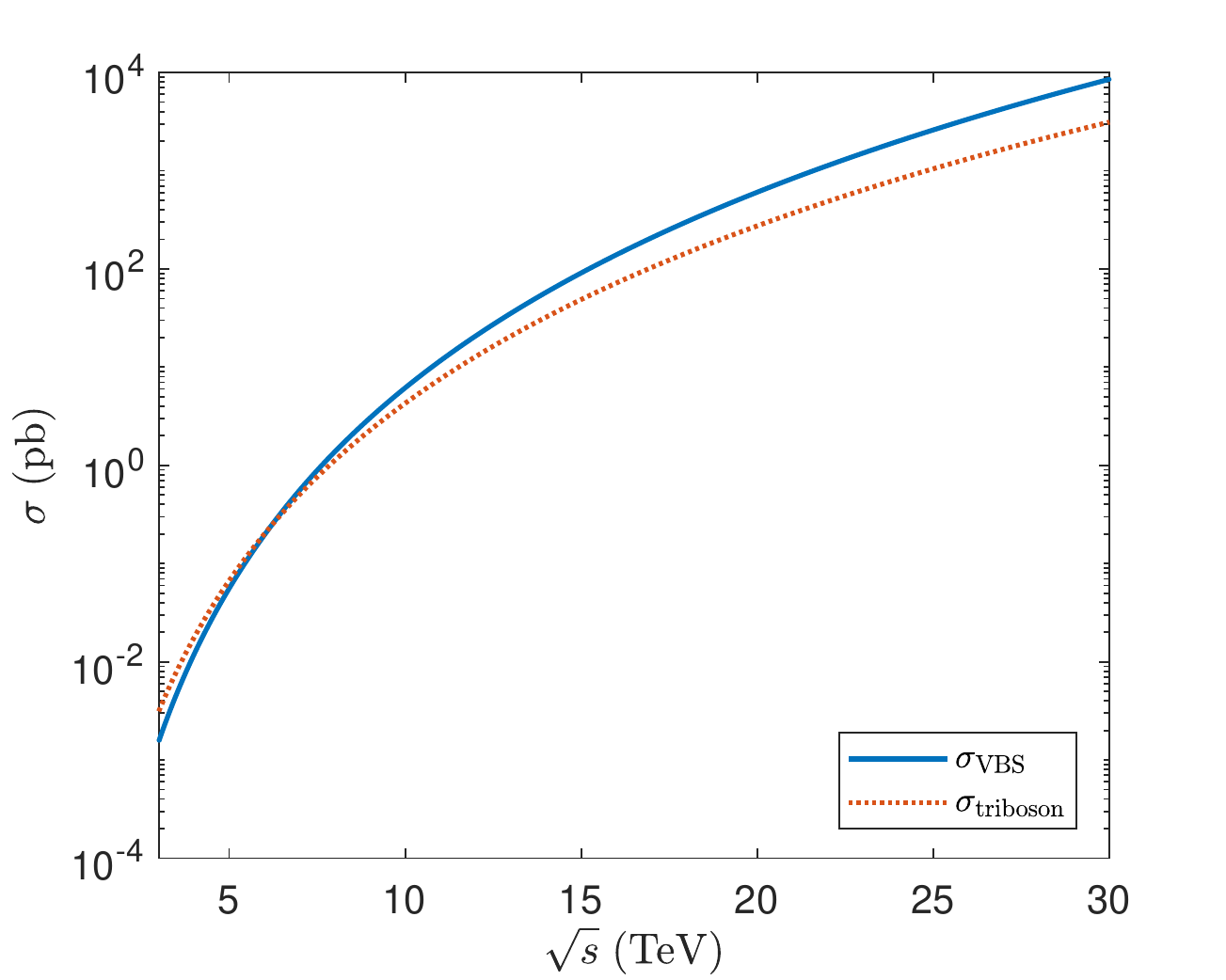}
\includegraphics[width=0.48\textwidth]{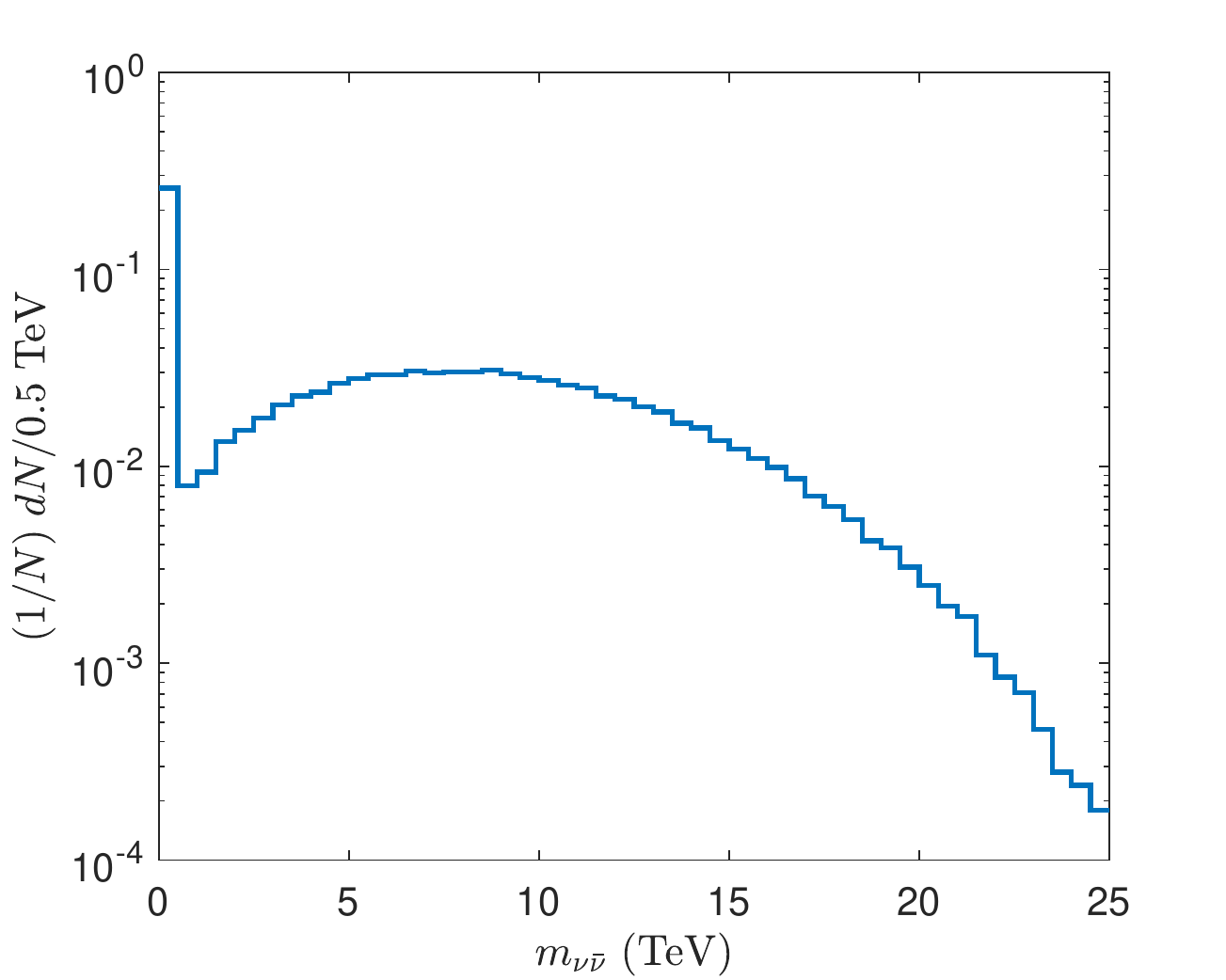}
\caption{\label{fig:llvvaa}Left: A comparison of $\sigma _{\rm VBS}$ and $\sigma _{\rm triboson}$ in Eq.~(\ref{eq.2.2.3}) for $\mu^+\mu^-\to \gamma\gamma\nu\bar{\nu}$, as a function of $\sqrt{s}$ with $f_{T_5}/\Lambda^4=0.5\;{\rm TeV}^{-4}$ allowed by LHC constraints. Right: The normalized distribution of $m_{\nu\bar{\nu}}$ for the aQGCs contribution in the process $\mu^+\mu^-\to \gamma\gamma \nu \bar{\nu}$ with $\sqrt{s}=30$ TeV.
}
\end{center}
\end{figure}

To demonstrate that the tri-boson processes can archive competitive sensitivity to aQGCs at the muon collider, we consider the process $\mu^+\mu^-\to \gamma\gamma \nu \bar{\nu}$ for illustration below.
The typical Feynman diagrams of the aQGC contribution to this process are shown in Fig.~\ref{fig:llvvaadiagram}.
For illustration, we only involve $O_{T_5}$ as an example in which both VBS and tri-boson contributions are present.
The VBS contribution in Fig.~\ref{fig:llvvaadiagram} (a) is calculated using effective vector boson approximation~\cite{eva1,eva2,eva3}.
\begin{equation}
\begin{split}
&\sigma_{\rm VBS} (\mu^+\mu^-\to \bar{\nu}\nu \gamma\gamma)=\sum _{\lambda _1\lambda _2\lambda _3\lambda _4}\int d\xi _1\int d\xi_2 f_{W_{\lambda _1}^-/\mu^-}(\xi _1)f_{W_{\lambda _2}^+/\mu^+}(\xi _2)\sigma _{W_{\lambda _1}^+W_{\lambda _2}^-\to \gamma_{\lambda _3}\gamma_{\lambda _4}}(\hat{s}),\\
&f_{W_{+1}^-/\mu _L^-}(\xi)=f_{W_{-1}^+/\mu_L^+}(\xi)=\frac{e^2}{8\pi^2s_W^2}\frac{(1-\xi)^2}{2\xi}\log \frac{\mu_f^2}{M_W},\\
&f_{W_{-1}^-/\mu _L^-}(\xi)=f_{W_{+1}^+/\mu_L^+}(\xi)=\frac{e^2}{8\pi^2s_W^2}\frac{1}{2\xi}\log \frac{\mu_f^2}{M_W},\\
&f_{W_{0}^-/\mu _L^-}(\xi)=f_{W_{0}^+/\mu_L^+}(\xi)=\frac{e^2}{8\pi^2s_W^2}\frac{1-\xi}{\xi},\\
&f_{W_{\lambda}^{\pm}/\mu _R^{\pm}}=0,\;\;\;\;f_{W_{\lambda}^{\pm}/\mu _R^{\pm}}=\frac{f_{W_{\lambda}^{\pm}/\mu _L^{\pm}}+f_{W_{\lambda}^{\pm}/\mu _R^{\pm}}}{2},\\
\end{split}
\label{eq.2.2.2}
\end{equation}
where $s_W$ is the sine of the weak mixing angle $\theta _W$, $\sqrt{\hat{s}}=\sqrt{\xi _1\xi _2 s}$ is the center of mass~(c.m.) energy of $W^+W^-\to\gamma\gamma$ with $\sqrt{s}$ being the c.m. energy of $\mu^+\mu^-$, and  $\mu _f$ is the factorization scale set to be $\sqrt{\hat{s}}/4$~\cite{eva3}.
The cross-section $\sigma _{\rm triboson}$ of the tri-boson diagram in Fig.~\ref{fig:llvvaadiagram} (b) is estimated as $\sigma (\mu^+\mu^-\to Z \gamma\gamma) \times {\rm Br}(Z\to \nu\bar{\nu})$, and $ {\rm Br}(Z\to \nu\bar{\nu})$ is taken as $20\%$~\cite{pdg} in the numerical calculation.
At the leading order of $s$, the analytical cross-sections are given by
\begin{equation}
\begin{split}
&\sigma_{\rm VBS}=\frac{e^4 f_{T_5}^2 s^3 \left(1-s_W^2\right)^2 \left[20 \log \left(\frac{s}{16 M_W^2}\right) \left(30 \log \left(\frac{s}{16 M_W^2}\right)-67\right)+943\right]}{110592000 \pi ^5 \Lambda ^8 s_W^4}\;,\\
&\sigma_{\rm triboson}=\frac{e^2 f_{T_5}^2 s^3 \left(48 s_W^8-64 s_W^6+40 s_W^4-12 s_W^2+3\right)}{138240 \pi ^3 \Lambda ^8 s_W^2 \left(s_W^2-1\right)}\times {\rm Br}(Z\to \nu\bar{\nu})\;.\\
\end{split}
\label{eq.2.2.3}
\end{equation}
The numerical results can be obtained by using the \verb"MadGraph5_aMC@NLO" toolkit~\cite{madgraph,feynrules}.
The cross-sections as a function of c.m. energy $\sqrt{s}$ with $f_{T_5}/\Lambda^4=0.5\;{\rm TeV}^{-4}$ are shown in the left panel of Fig.~\ref{fig:llvvaa}. Although the VBS process overwhelms for larger $\sqrt{s}$, one can see that the tri-boson process is of the same order of amplitude as the VBS process. This is because the momentum dependence in the Feynman rule of the $Z\gamma\gamma\gamma$ or $ZZ\gamma\gamma$ vertex cancels the $1/s$ in the propagator of tri-boson process.
It turns out that the tri-boson process is competitive with the VBS processes to study the aQGCs for $\sqrt{s}<30\;{\rm TeV}$.

Before the detector simulation, in the right panel of Fig.~\ref{fig:llvvaa}, we also show the normalized distribution of the invariant mass of neutrinos~(denoted as $m_{\nu\bar{\nu}}$) at $\sqrt{s}=30\;{\rm TeV}$. The tri-boson process exhibits a peak around $m_{\nu\bar{\nu}}\sim M_Z$ due to the onshell $Z$ boson decay.
Using $m_{\nu\bar{\nu}}$ as a mild cut, we obtain $\sigma_{O_{T_5}}(m_{\nu\bar{\nu}} > 100\;{\rm GeV})=9048.0\;{\rm pb}$ and $\sigma_{O_{T_5}}(m_{\nu\bar{\nu}} < 100\;{\rm GeV})=2797.7\;{\rm pb}$~(Eq.~(\ref{eq.2.2.3}) yields $\sigma _{\rm VBS}=8534.4\;{\rm pb}$ and $\sigma _{\rm triboson}=3128.4\;{\rm pb}$).
In principle, the cross section $\sigma_{O_{T_5}}$ obtained in the Monte-Carlo simulation should be the sum of $\sigma_{\rm VBS}$, $\sigma_{\rm triboson}$ as well as their interference. It does not include the SM contribution. $\sigma_{\rm VBS}$ and $\sigma_{\rm triboson}$ are analytically calculated using Eq.~(\ref{eq.2.2.3}). The numbers demonstrate that the cut $m_{\nu\bar{\nu}}>(<)~100$ GeV would keep most of VBS (tri-boson) contribution. The small discrepancy should be attributed to the effective vector approximation for VBS and the kinematic cuts in the Monte-Carlo simulation.
Despite this discrimination, there exists inevitable interference between the two kinds of processes.

\subsection{\label{sec2.3}The contribution of aQGCs to the tri-photon process}

\begin{figure}
\begin{center}
\includegraphics[width=0.7\textwidth]{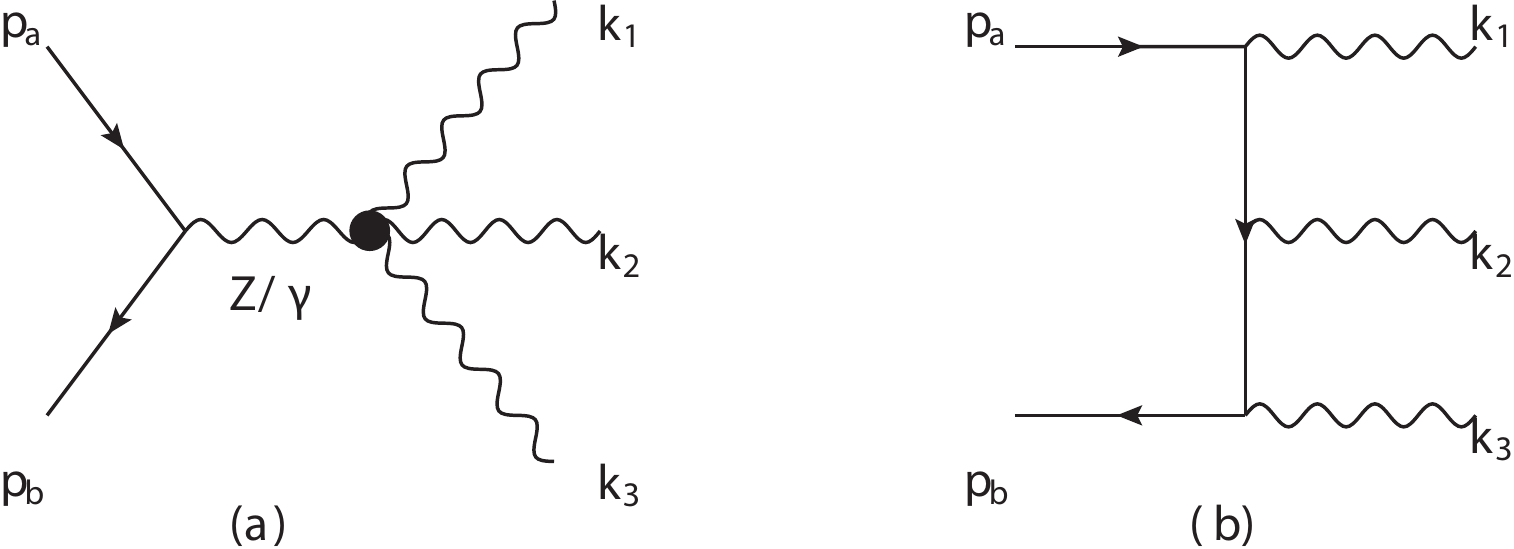}
\caption{\label{fig:diagram}Feynman diagrams for the process $\mu^+\mu^-\to \gamma\gamma\gamma$. The diagrams induced by $O_{T_i}$ operators are shown in the left panel, and a representative diagram in the SM is shown in the right panel. In the SM, there are other five diagrams by permutating the photons in the final state.}
\end{center}
\end{figure}

In fact, most of signal channels induced by aQGCs are composed of both VBS and tri-boson production topologies with the decay of massive EW gauge bosons.
However, unlike the process $\mu^+\mu^-\to \gamma\gamma\nu\bar{\nu}$, it is interesting to notice that the tri-photon production $\mu^+\mu^-\to \gamma\gamma\gamma$ has only one topology induced by one single electroweak vertex. Moreover, the photons in final states have no subsequent decays and thus the production rate would be enhanced compared with the production of massive EW gauge bosons. As a result, the tri-photon process provides a distinctive channel for the study of anomalous $O_{T_i}$ operators. We analyze this tri-photon channel in this subsection.

At tree level, there are only two Feynman diagrams ($\gamma$ or $Z$ mediated) induced by $O_{T_i}$ as shown in Fig.~\ref{fig:diagram}~(a).
The SM background is from the process $\mu^+\mu^-\to \gamma\gamma \gamma$.
There are six Feynman diagrams, one of which is shown in Fig.~\ref{fig:diagram}~(b). The other five diagrams can be obtained by permutating the photons in the final states.

The total cross-section with aQGCs can be written as
\begin{equation}
\begin{split}
&\sigma_{\rm aQGC}(f_{T_i}) = \sigma _{\rm SM}+\sigma _{O_{T_i}}(f_{T_i})+\sigma _{\rm int}(f_{T_i})\;,
\end{split}
\label{eq.2.4}
\end{equation}
where $\sigma _{\rm SM}$, $\sigma _{O_{T_i}}$ and $\sigma _{\rm int}$ are the SM cross-section, the $O_{T_i}$ operator induced cross-section and their interference, respectively.
For a large $\sqrt{s}$, one can ignore the masses of muon and $Z$ boson.
The analytical results of $\sigma _{O_{T_i}}$ and $\sigma _{\rm int}$ are given by
\begin{equation}
\begin{split}
&\sigma _{\rm int}=\frac{e^4 s (384 \log (2)-215) \left( (1-4 s_W^2) (4 \alpha _1+3 \alpha _2)+16 c_W s_W (4 \alpha _3 +3 \alpha _4)\right)}{110592 \pi ^3 \Lambda ^4 c_W s_W},\\
&\sigma _{O_{T_i}}=\frac{e^2 s^3 }{276480 \pi ^3 \Lambda ^8 c_W^2 s_W^2}\left(8 c_W s_W(1-4s_W^2) (16 \alpha _1 \alpha _3+7 \alpha _1 \alpha _4+7 \alpha _2 \alpha _3+4 \alpha _2 \alpha _4)\right.\\
&\left.+(1-4s_W^2+8s_W^4) \left(8 \alpha _1^2+7 \alpha _1 \alpha _2+2 \alpha _2^2\right)+ 128 c_W^2 s_W^2\left(8 \alpha _3^2+7 \alpha _3 \alpha _4+2 \alpha _4^2\right)\right),
\end{split}
\label{eq.2.5}
\end{equation}
with
\begin{equation}
\begin{split}
&\alpha _1 = c_W^3 s_W (f_{T_5}+f_{T_6}-4 f_{T_8})+c_W s_W^3 (f_{T_0}+f_{T_1}-f_{T_5}-f_{T_6}),\\
&\alpha _2 = c_W^3 s_W (f_{T_7}-4 f_{T_9})+c_W s_W^3 (f_{T_2}-f_{T_7}),\\
&\alpha _3 = c_W^4 f_{T_8}+\frac{1}{2} c_W^2 s_W^2 (f_{T_5}+f_{T_6})+\frac{1}{4} s_W^4 (f_{T_0}+f_{T_1}),\\
&\alpha _4 = c_W^4 f_{T_9}+\frac{1}{2} c_W^2 f_{T_7} s_W^2+\frac{f_{T_2} s_W^4}{4}.\\
\end{split}
\label{eq.2.6}
\end{equation}
Here $c_W$ denotes $\cos(\theta_W)$.
By using \verb"MadGraph5@NLO", the results of $\sigma _{\rm SM}$ are listed in Table~\ref{tab.sigmasm}.
They are obtained with default basic cuts for photons
\begin{equation}
\begin{split}
&p_{T,\gamma}>10\;{\rm GeV},\;\;|\eta _{\gamma}|<2.5,\;\;\Delta R_{\gamma \gamma}>0.4,\\
\end{split}
\label{eq.2.7}
\end{equation}
where $p_{T,\gamma}$ is the transverse momentum of the photon, $\eta _{\gamma}$ is the pseudo-rapidity, and $\Delta R=\sqrt{\Delta \phi^2+\Delta \eta^2}$ with $\Delta \phi$ and $\Delta \eta$ being the difference of azimuth angles and pseudo-rapidities of the photons.

The tri-photon process is particularly more sensitive to $O_{T_{8,9}}$ operators which only contribute to neutral aQGCs.
For illustration, taking $O_{T_9}$ and assuming non-vanishing $f_{T_9}/\Lambda^4=0.92\; {\rm TeV}^{-4}$, we obtain $\sigma _{O_{T_9}}=0.26\;{\rm pb}$ which is at least $40$ times larger than $\sigma _{\rm SM}$ at $\sqrt{s}=3\;{\rm TeV}$.
Meanwhile, $\sigma _{\rm int}=0.013\;{\rm pb}$. Note that $\sigma _{\rm int}\propto f_{T_i}$, if the constraints can be at least two orders of magnitude severer, the interference terms become dominant.

At large $s$, the validity of the SMEFT should be taken into account.
Unlike the case of VBS, the amplitude of the triphoton induced by aQGCs is suppressed by a propagator for large $s$.
A similar case was studied for the diboson production induced by dimension-8 operators contributing to neutral triple gauge couplings~(nTGCs) at an $e^+e^-$ collider through the $s$-channel~\cite{ntgc7}.
It has been found that the unitarity constraint is only relevant at very low luminosity.
Therefore, we omit the unitarity issue in this work.

\begin{table}
\begin{center}
\begin{tabular}{c|c|c|c|c}
\hline
  & $3\;{\rm TeV}$ & $10\;{\rm TeV}$ & $14\;{\rm TeV}$ & $30\;{\rm TeV}$ \\
\hline
  $\sigma _{\rm SM}\;(\rm fb)$ & $5.96$ & $0.707$ & $0.383$ & $0.0953$ \\
\hline
\end{tabular}
\end{center}
\caption{\label{tab.sigmasm}The results of $\sigma _{\rm SM}$ for $\mu^+\mu^-\to \gamma\gamma\gamma$ after the basic cuts in Eq.~(\ref{eq.2.7}).}
\end{table}

\section{\label{sec3}Numerical results}

\subsection{\label{sec3.1}Event selection strategy}

\begin{figure}
\begin{center}
\includegraphics[width=0.6\textwidth]{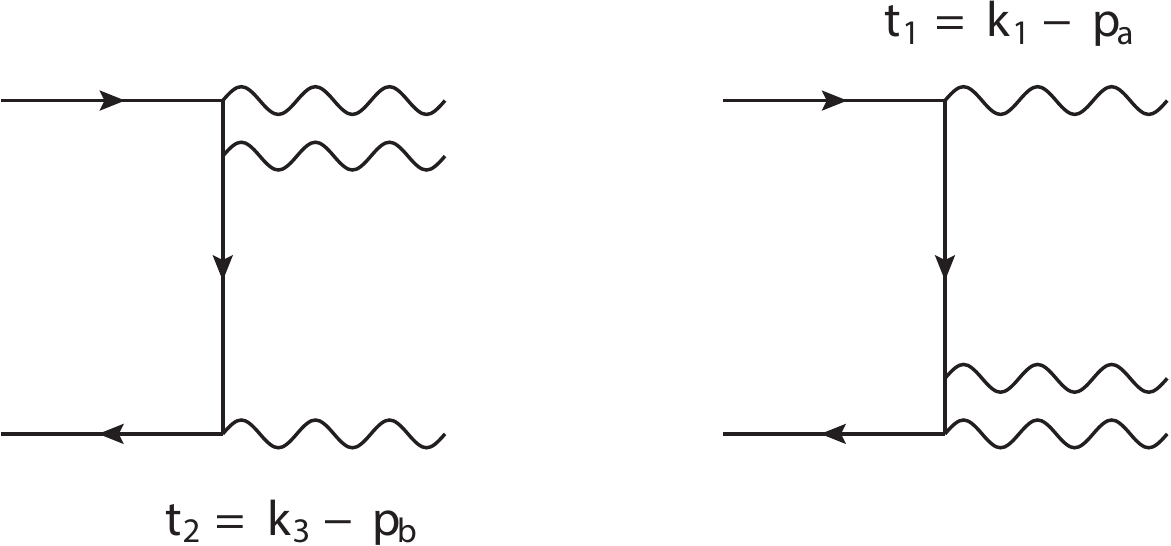}
\caption{\label{fig:t1t2}The SM diagram in Fig.~\ref{fig:diagram}~(b) can be treated as t-channel $2\to 2$ diagrams. }
\end{center}
\end{figure}

We use the \verb"MadGraph5_aMC@NLO" toolkit to perform the numerical calculation. The basic cuts in Eq.~(\ref{eq.2.7}) are also applied at partonic level.
A fast detector simulation is then performed by \verb"Delphes"~\cite{delphes} with the muon collider card.
The analyses of the signal and the background are completed by \verb"MLAnalysis"~\cite{Guo:2023nfu}.  
To study the kinematic features of the signal and background, the signal events are generated by assuming one operator at a time and the largest coefficients allowed by LHC constraints in Table~\ref{tab.1} are adopted.
We require at least three photons in the following studies.
All the numerical results are presented after this photon number cut~(denoted as $N_{\gamma}$ cut).

\begin{figure}
\begin{center}
\includegraphics[width=0.48\textwidth]{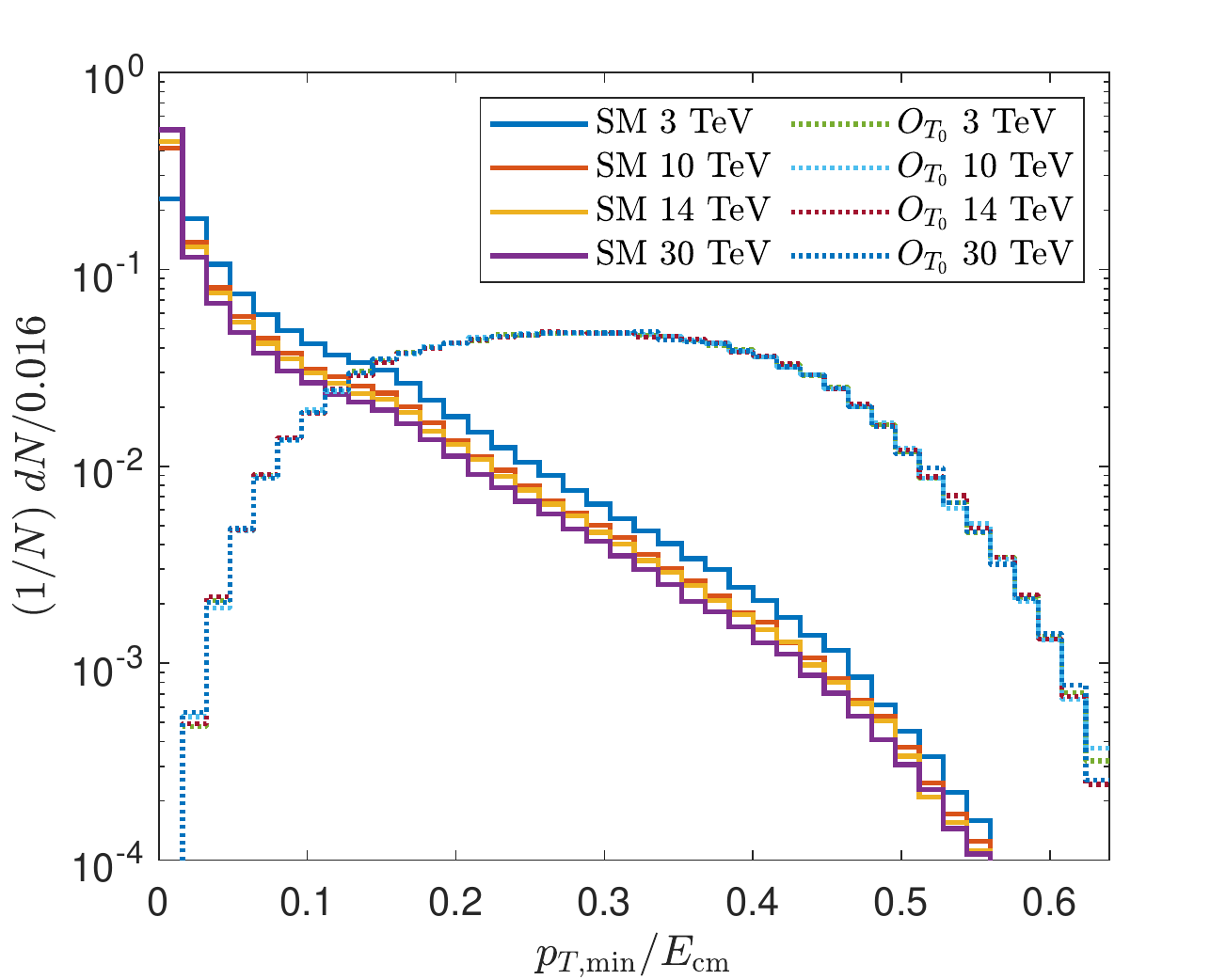}
\includegraphics[width=0.48\textwidth]{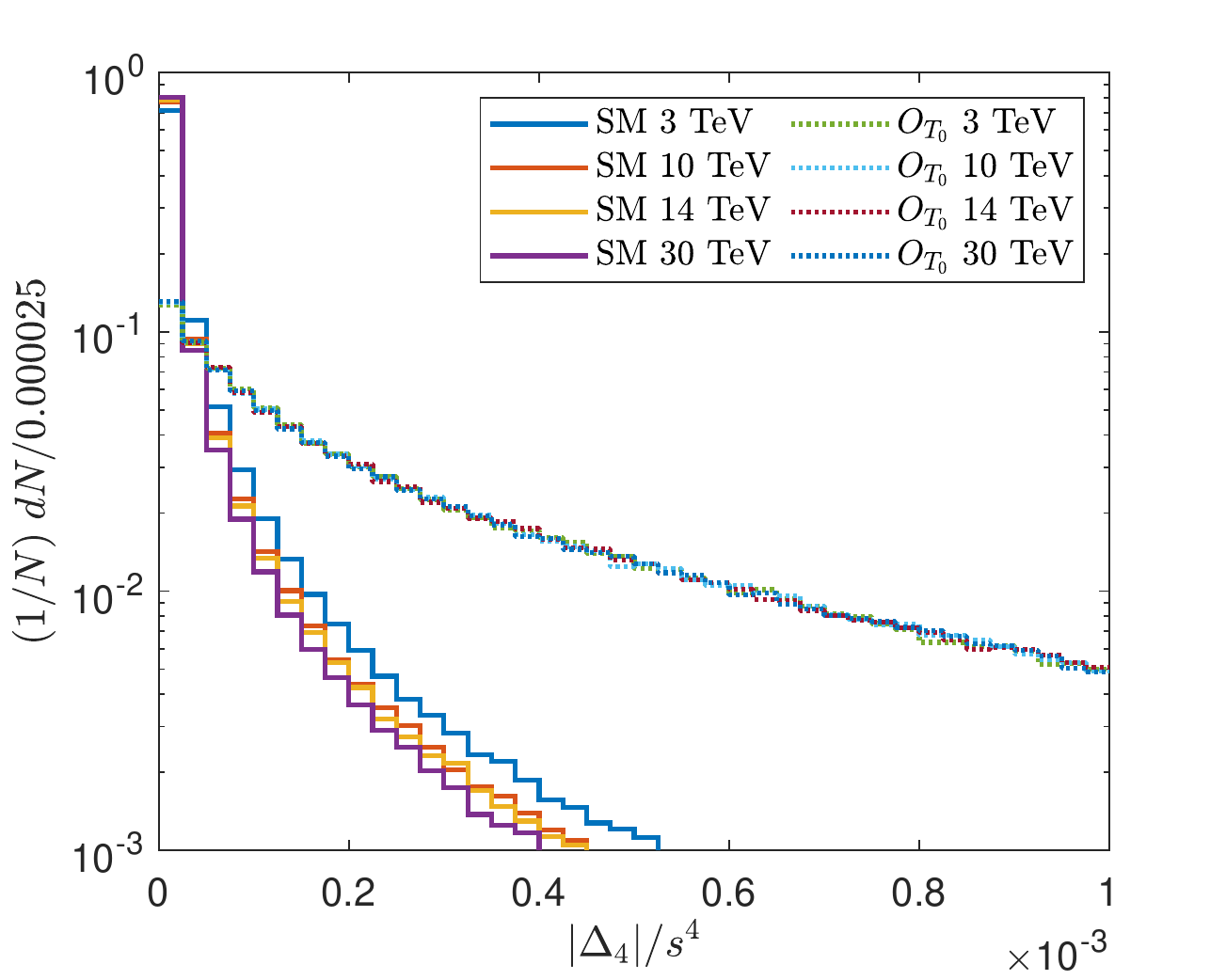}
\caption{\label{fig:feature}The normalized distributions of $p_{T,\rm min}/E_{\rm cm}$~(left panel) and $|\Delta _4|/s^4$~(right panel) for SM and $O_{T_i}$ after the $N_{\gamma}$ cut. }
\end{center}
\end{figure}
The aQGC diagrams in Fig.~\ref{fig:diagram}~(a) are through s-channel.
For the SM diagram in Fig.~\ref{fig:diagram}~(b), one can combine two photons as a virtual particle to form $2\to 2$ t-channel diagrams as Fig.~\ref{fig:t1t2} so that $\mathcal{A}\propto 1/t_1$ or $\mathcal{A}\propto 1/t_2$, where $t_1=(p_a-k_1)^2$, $t_2=(p_b-k_3)^2$ with $p_{a,b}$ and $k_{1,2,3}$ shown in Fig.~\ref{fig:diagram}. The other five diagrams can be transformed similarly. By contrast, the aQGC diagrams can only transform to s-channel $2\to 2$ diagrams.
We expect that the distribution of SM background events is dominantly around small $t$ and thus the transverse momentum of one of the photons is small.
Denoting $p_{T,\rm min}$ as the smallest transverse momentum of the three hardest photons and $E_{\rm cm}$ as the c.m. energy, the normalized distributions of $p_{T,\rm min}/E_{\rm cm}$ for the signal of $O_{T_0}$ and the background are displayed in the left panel of Fig.~\ref{fig:feature}.
The distributions for other $O_{T_i}$ and different $\sqrt{s}$ are similar.
We thus apply hard $p_T$ cut to events, i.e., $p_{T,\rm min}/E_{\rm cm}>0.12$.

Another observable sensitive to IR divergence is the Gram determinant $\Delta_4$~\cite{Byckling} which appears in the denominator of the phase space integration, defined as
\begin{equation}
\begin{split}
&\Delta_4=\frac{-1}{16} \left(2 s_1 (s t_2 (s_2+t_1)+(s t_1-s_2t_2) (s_2-t_1))+(s (t_1-t_2)+s_2 t_2)^2+s_1^2 (s_2-t_1)^2\right)\;,\\
\end{split}
\label{eq.3.1}
\end{equation}
where $s_1=(k_1+k_2)^2$, $s_2=(k_2+k_3)^2$ with $p_{a,b}$ and $k_{1,2,3}$ shown in Fig.~\ref{fig:diagram}, and $t_{1,2}$ in Fig.~\ref{fig:t1t2}. The IR divergence shows up in the region $|\Delta _4|\to 0$, and thus one expects that the SM events sharply peak at small $|\Delta _4|$.
Since the permutation of photons in the final state is symmetric, in the calculation of $\Delta _4$, $k_1$ is assigned as the momentum of the hardest photon, $k_2$ as the second hardest, and $k_3$ as the third hardest.
The normalized distributions of $\left|\Delta _4\right|/s^4$ for the signals and the background are shown in the right panel of Fig.~\ref{fig:feature}.
We require the events satisfying $\left|\Delta _4\right|/s^4>3\times 10^{-5}$.

The cross-sections after event selection strategy and the cut efficiencies are shown in Table~\ref{tab.cutflow}.
Note that in the event selection strategy, $p_{T,\rm min}$ and $\Delta _4$ are rescaled to be dimensionless.
The cut efficiencies for the contributions of $O_{T_i}$ should be the same for different $\sqrt{s}$, which is also implied in the distributions shown in Fig.~\ref{fig:feature}.
Thus, only the cross-sections at $\sqrt{s}=3$ TeV are shown for $O_{T_i}$ in Table~\ref{tab.cutflow}. We find that the cut efficiency for our signal becomes $\sim 60\%$.
For the SM background, the cut efficiencies decrease with the growth of $\sqrt{s}$, and can be as small as $7.24\%$ at $\sqrt{s}=30\;{\rm TeV}$.

Another important background would be the beam induced background, such as the radiated photons tangent to electron trajectories inside the collider.
The $p_{T,\rm min}$ and $\Delta_4$ cuts would be helpful to remove such background. However, Delphes does not include the effects of this background contribution at this stage. A detailed quantitative analysis is deferred to future works.

\begin{table*}
\begin{center}
\begin{tabular}{c|c|ccc|c}
\hline
$\sqrt{s}$ &  & $N_{\gamma}$ cut & $p_{T,\rm min}/E_{\rm cm}>0.12$ & $|\Delta _4|/s^4>3\times 10^{-5}$ & efficiency $\epsilon$ \\
\hline
$3$ TeV  &    & $4.495$  & $1.085$  & $0.686$  & $0.115$  \\
$10$ TeV & SM & $0.533$  & $0.098$  & $0.062$  & $0.0877$ \\
$14$ TeV &    & $0.289$  & $0.050$  & $0.031$  & $0.0809$ \\
$30$ TeV &    & $0.0718$ & $0.0109$ & $0.0069$ & $0.0724$ \\
\hline
         & $O_{T_0}$ & $0.0125$ & $0.0117$ & $0.0103$ & $0.606$\\
         & $O_{T_2}$          & $0.0204$ & $0.0193$ & $0.0171$ & $0.624$\\
 $3$ TeV & $O_{T_5}$ & $5.746$  & $5.394$  & $4.720$  & $0.596$\\
         & $O_{T_7}$          & $4.706$  & $4.443$  & $3.935$  & $0.613$\\
         & $O_{T_8}$          & $163.72$ & $153.61$ & $134.16$ & $0.590$\\
         & $O_{T_9}$          & $189.36$ & $178.86$ & $158.14$ & $0.606$\\
\hline
\end{tabular}
\end{center}
\caption{\label{tab.cutflow}The cross-sections $\sigma_{\rm SM}$ and $\sigma_{O_{T_i}}$ (in unit of $\rm fb$) and cut efficiencies after selection strategy.
The results of aQGCs are obtained using the upper bound of the coefficients in Table~\ref{tab.1}.}
\end{table*}

\subsection{\label{sec3.2}Significance and the expected constraints}

\begin{figure*}
\begin{center}
\includegraphics[width=0.32\textwidth]{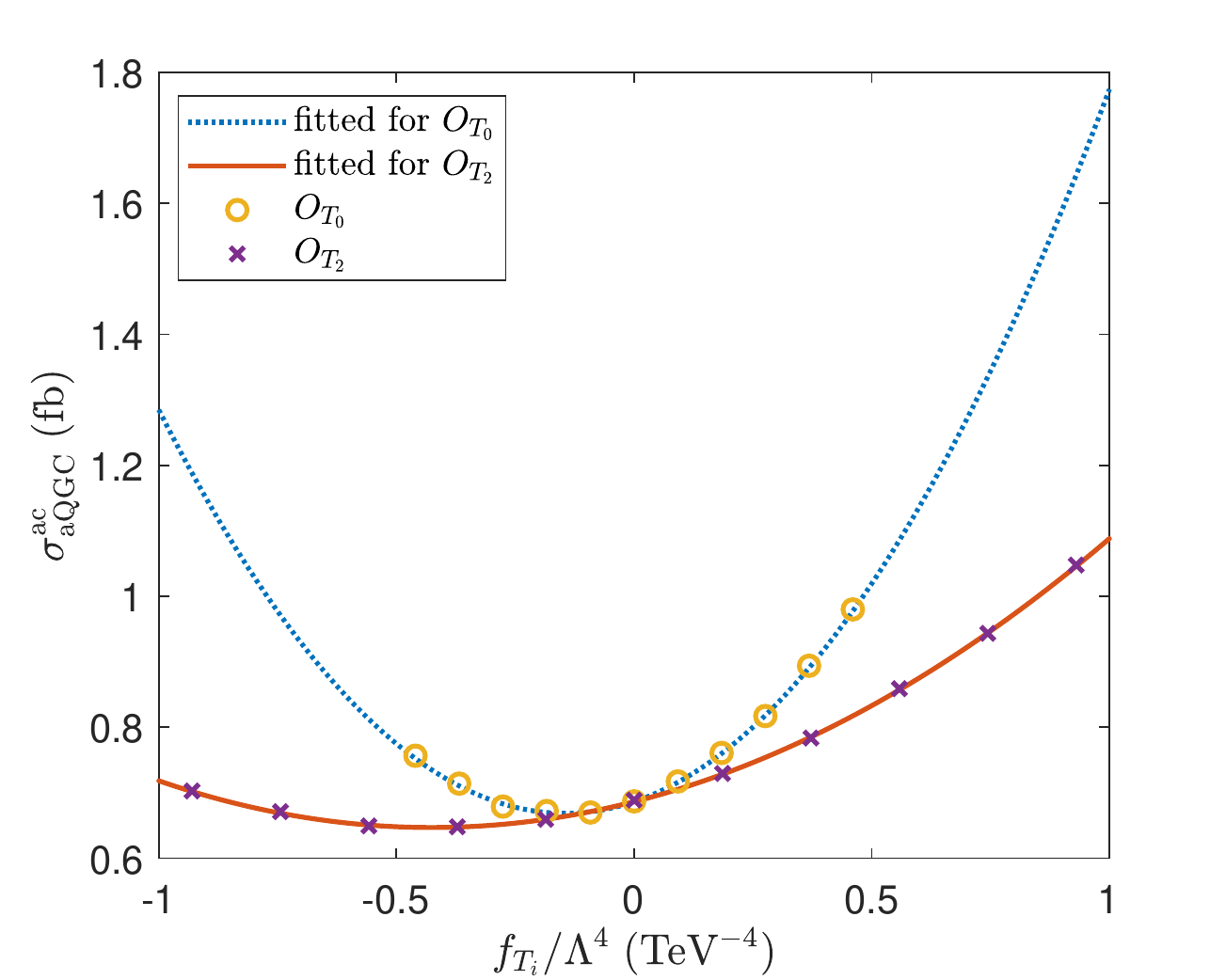}
\includegraphics[width=0.32\textwidth]{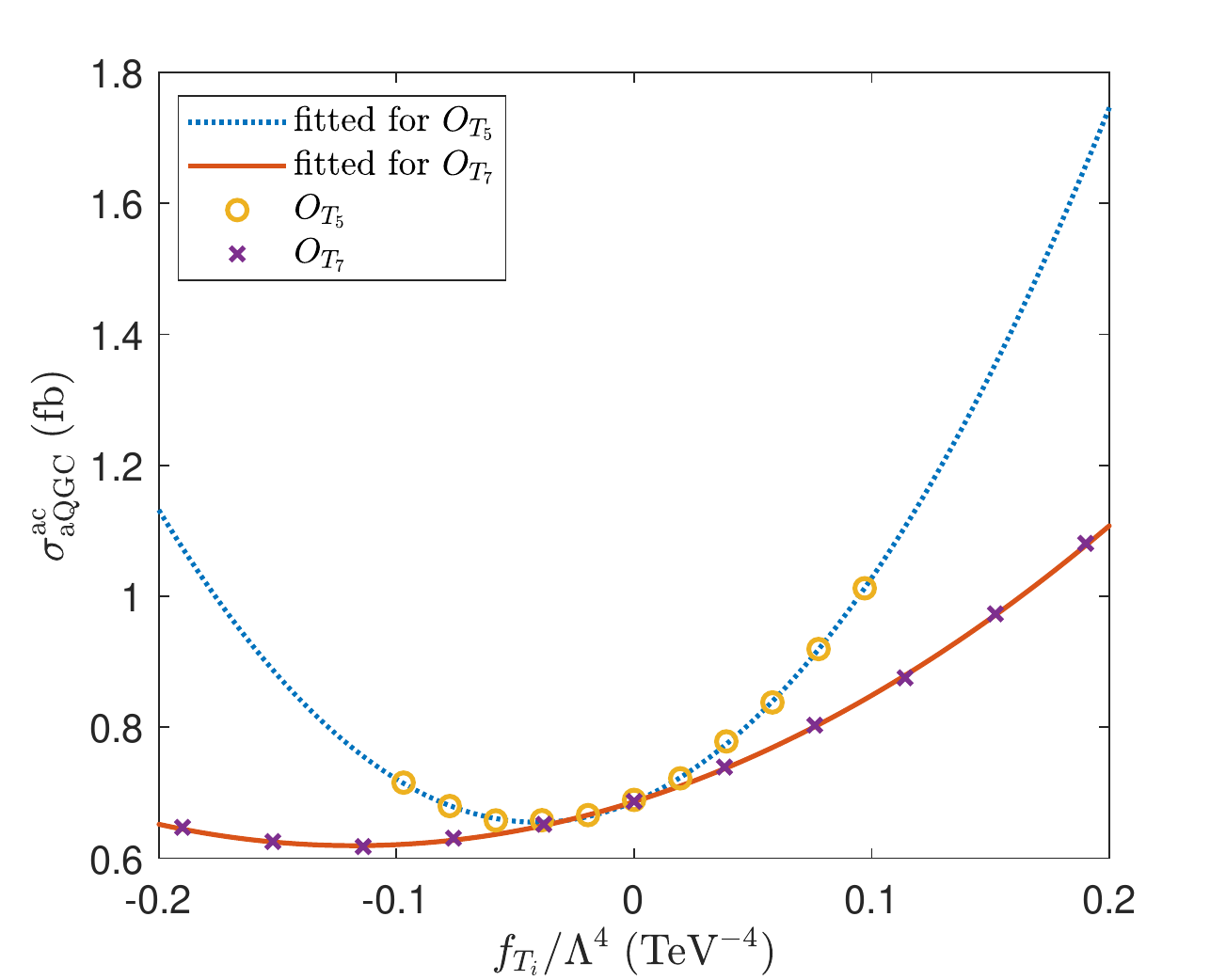}
\includegraphics[width=0.32\textwidth]{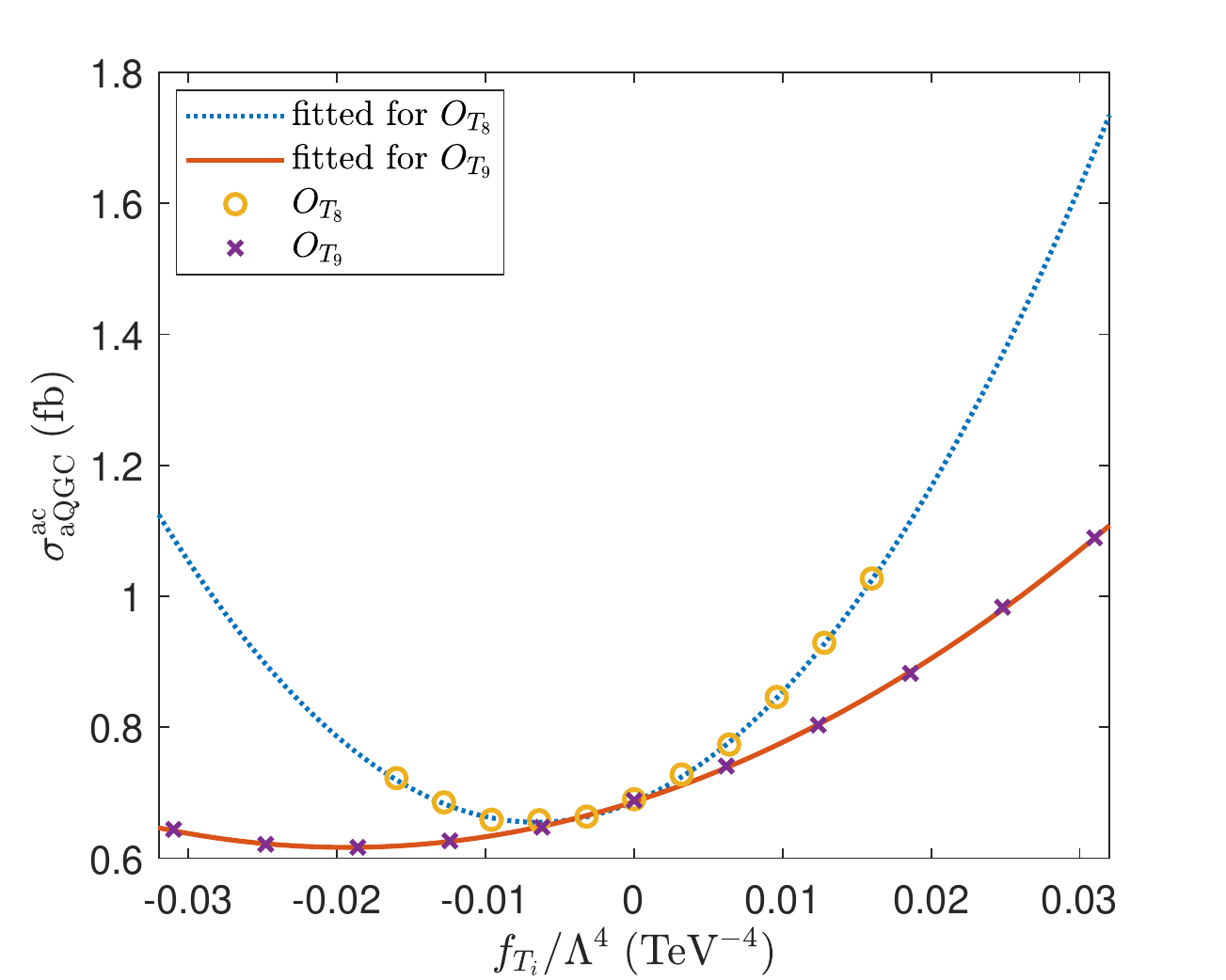}\\
\includegraphics[width=0.32\textwidth]{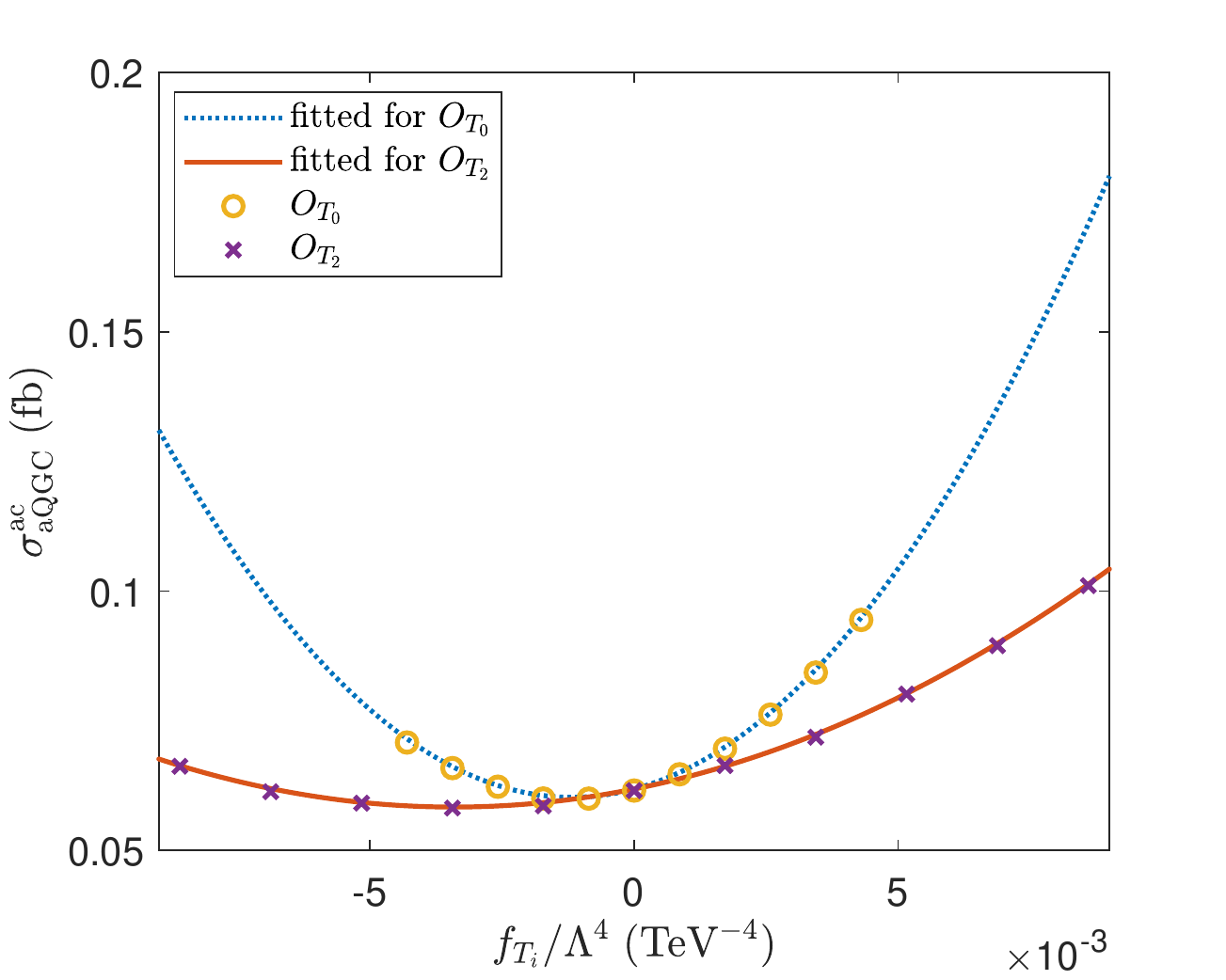}
\includegraphics[width=0.32\textwidth]{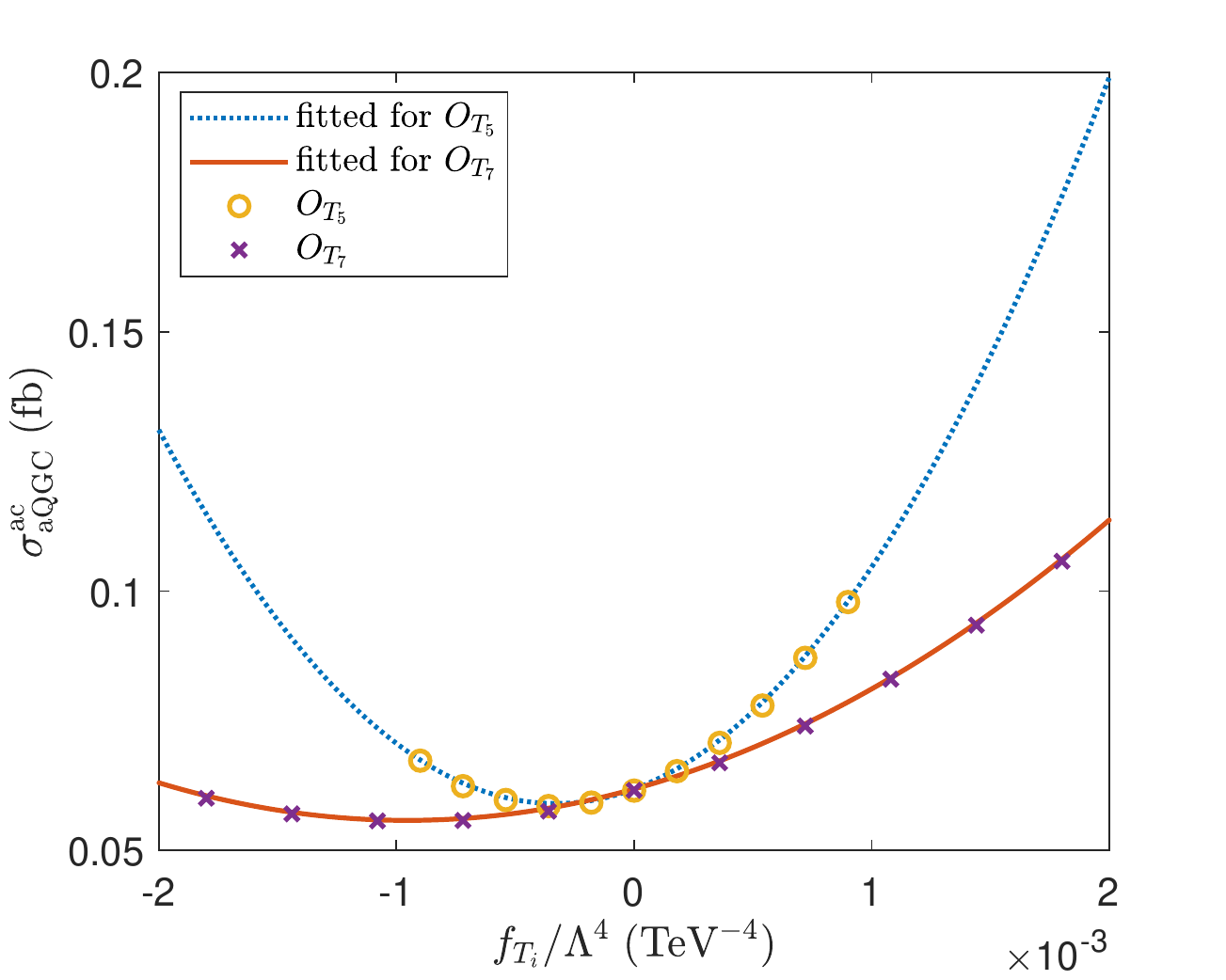}
\includegraphics[width=0.32\textwidth]{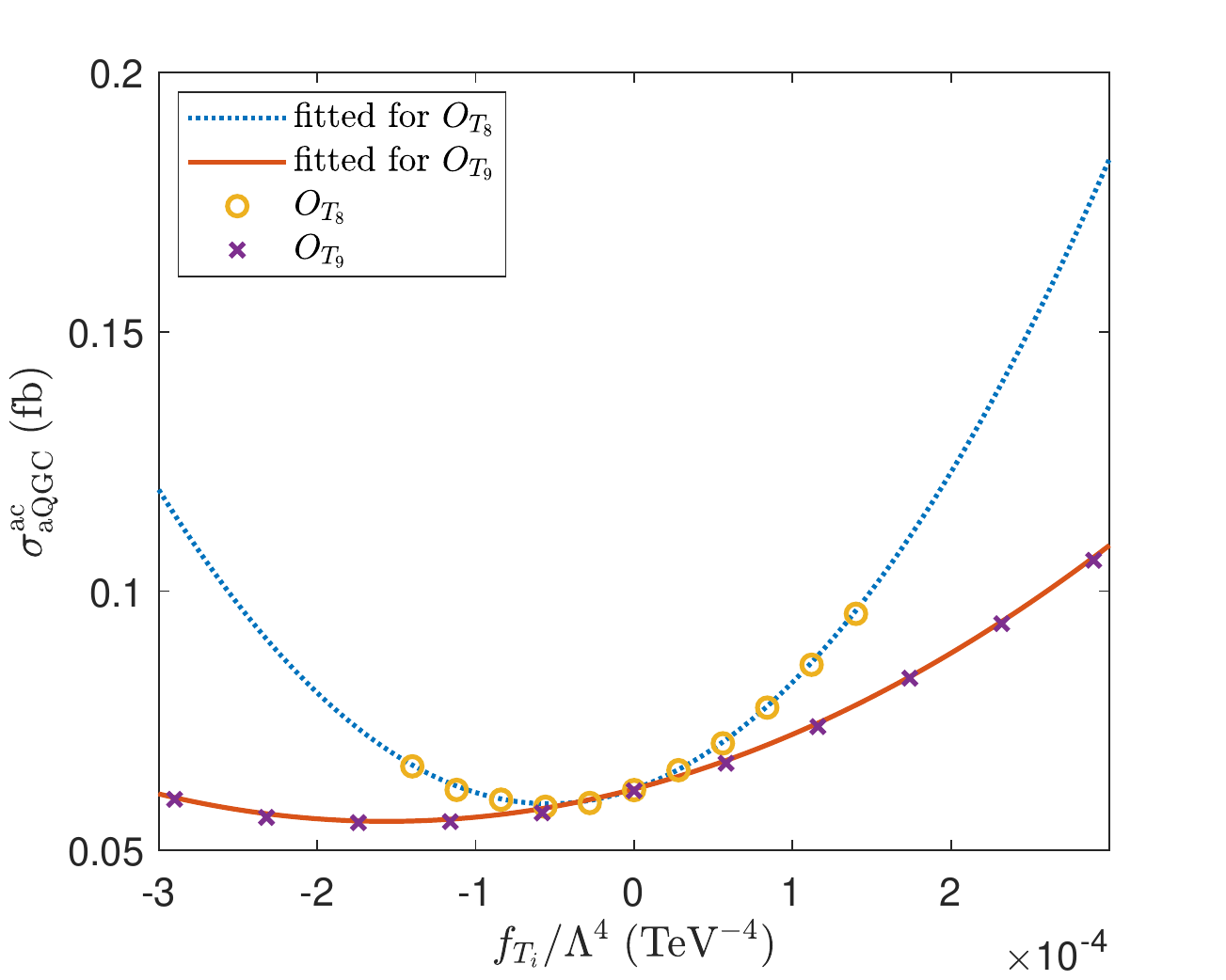}\\
\includegraphics[width=0.32\textwidth]{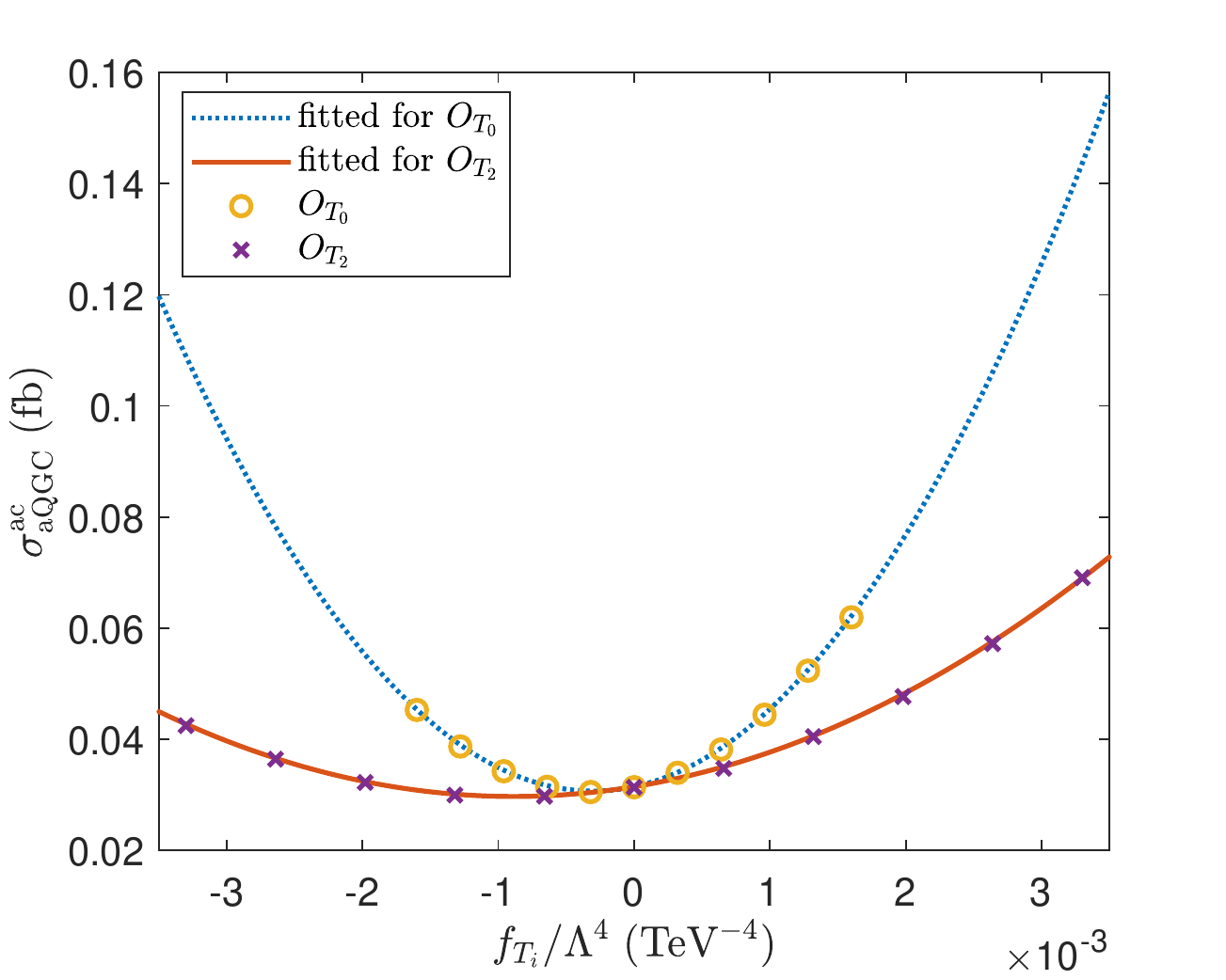}
\includegraphics[width=0.32\textwidth]{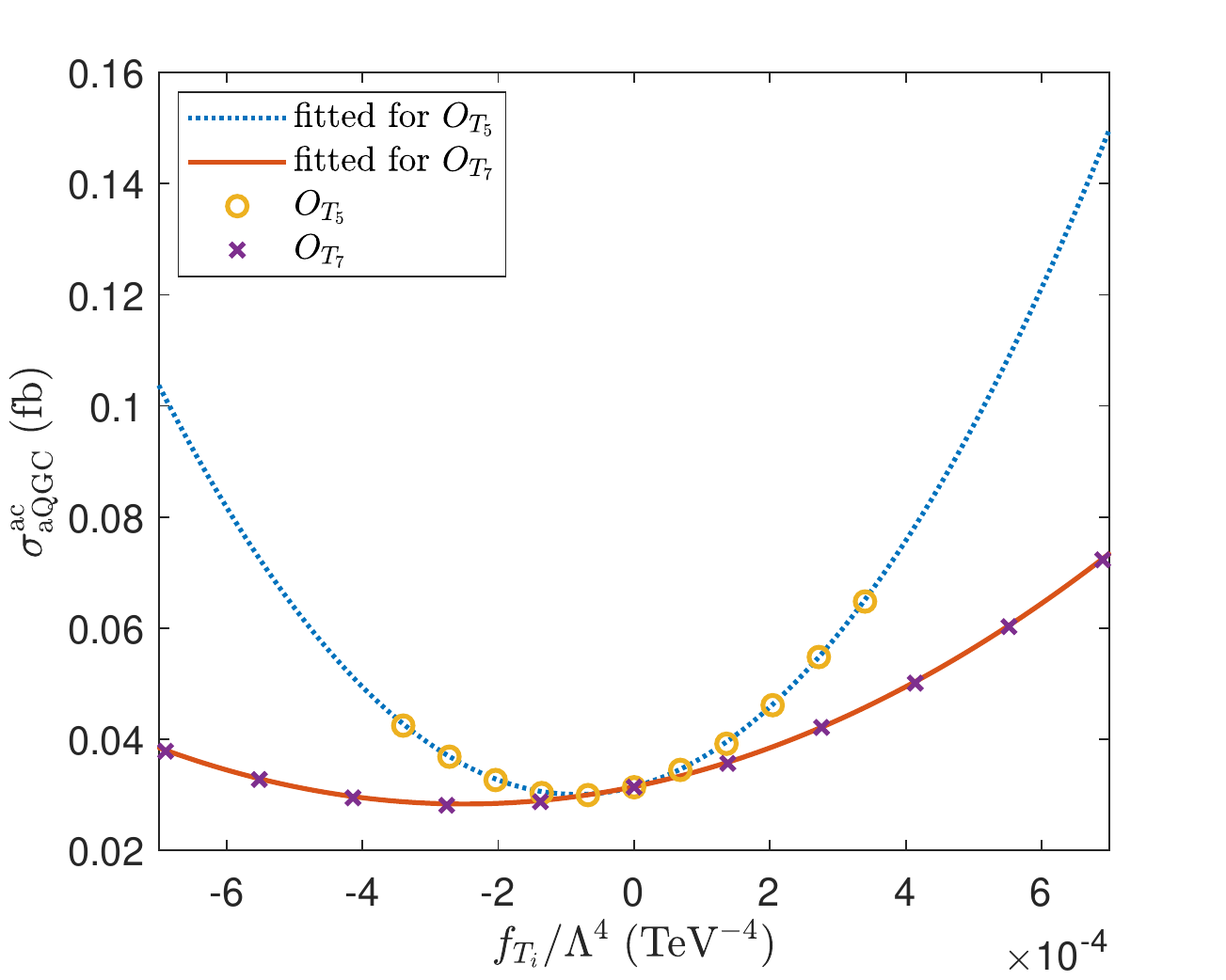}
\includegraphics[width=0.32\textwidth]{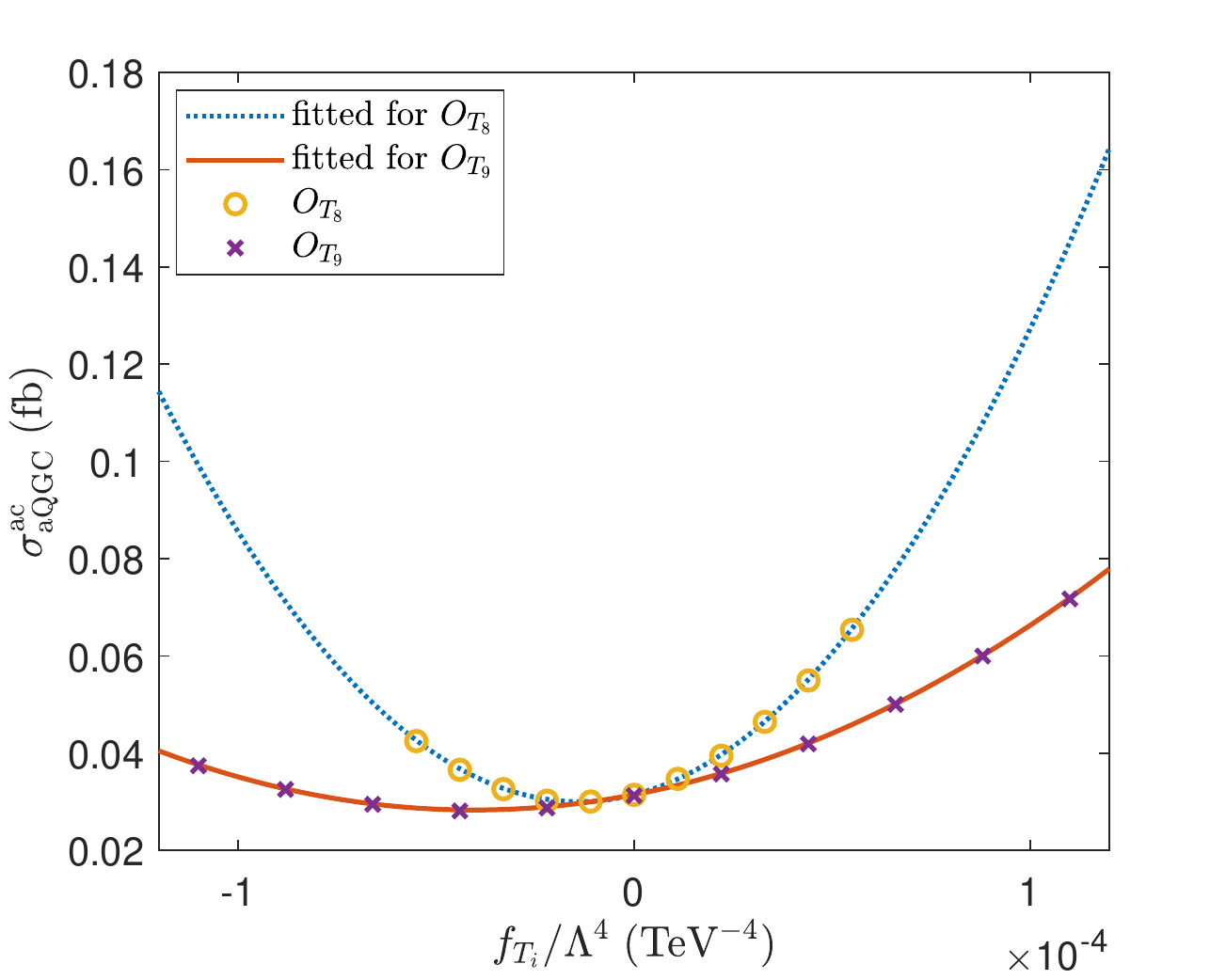}\\
\includegraphics[width=0.32\textwidth]{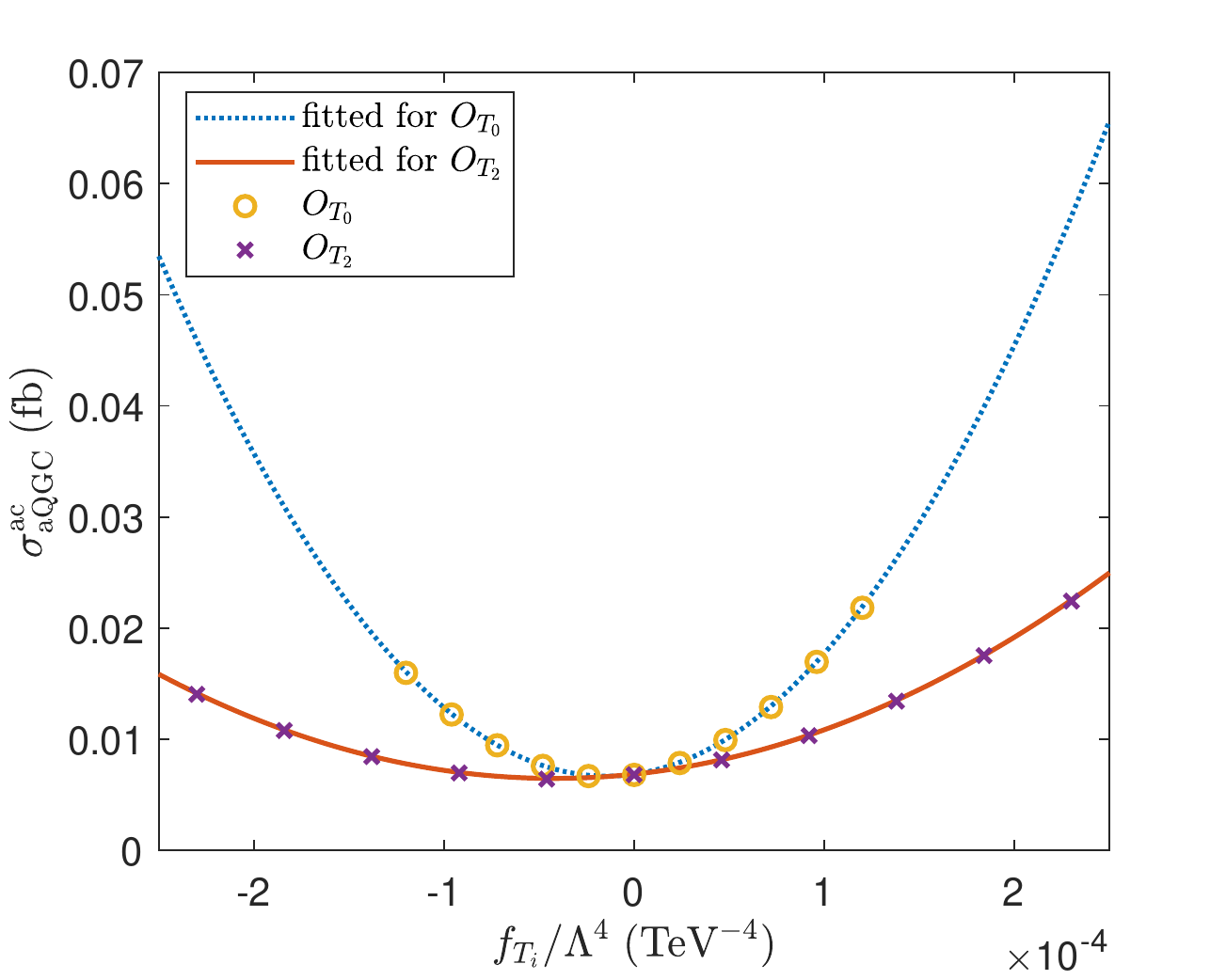}
\includegraphics[width=0.32\textwidth]{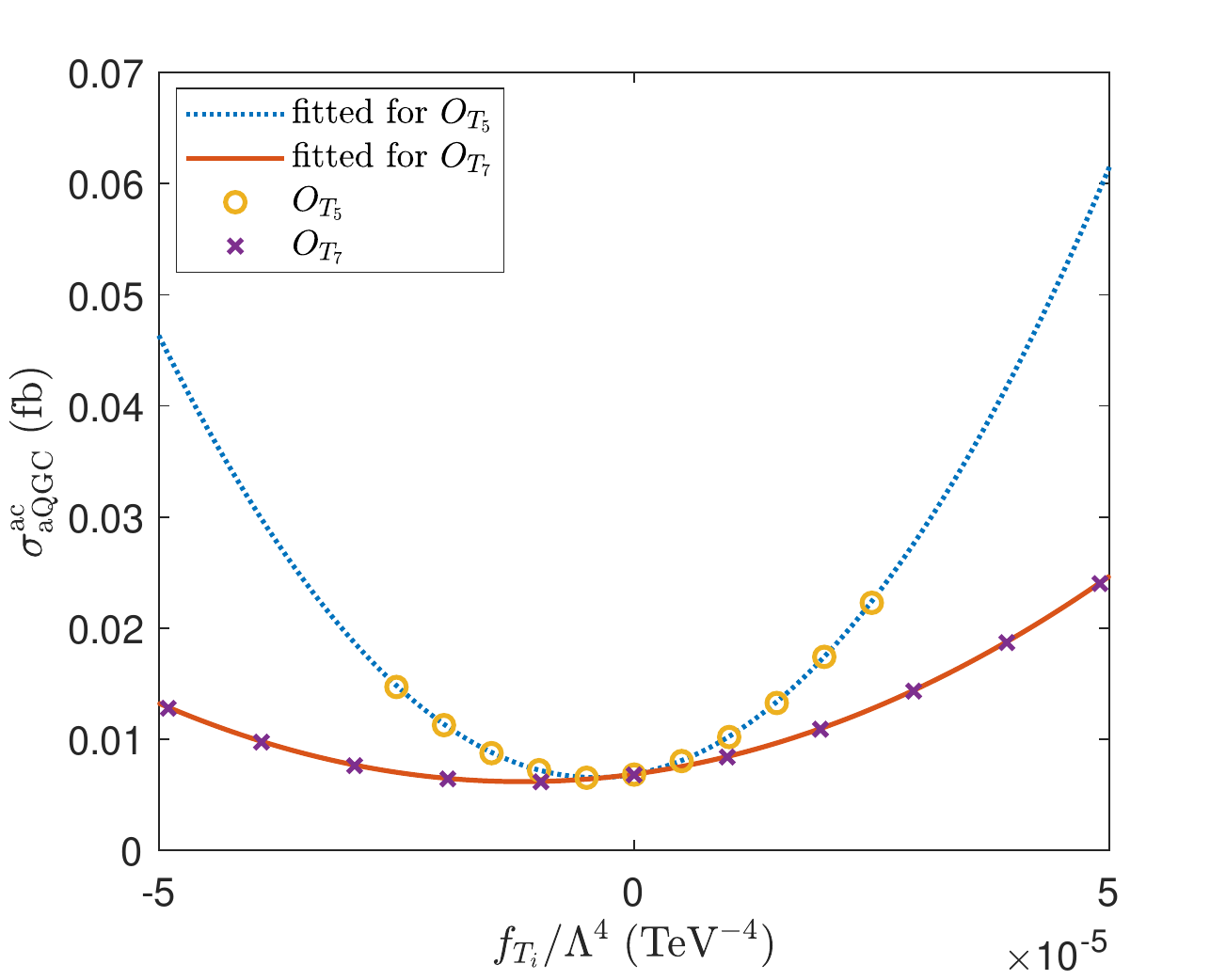}
\includegraphics[width=0.32\textwidth]{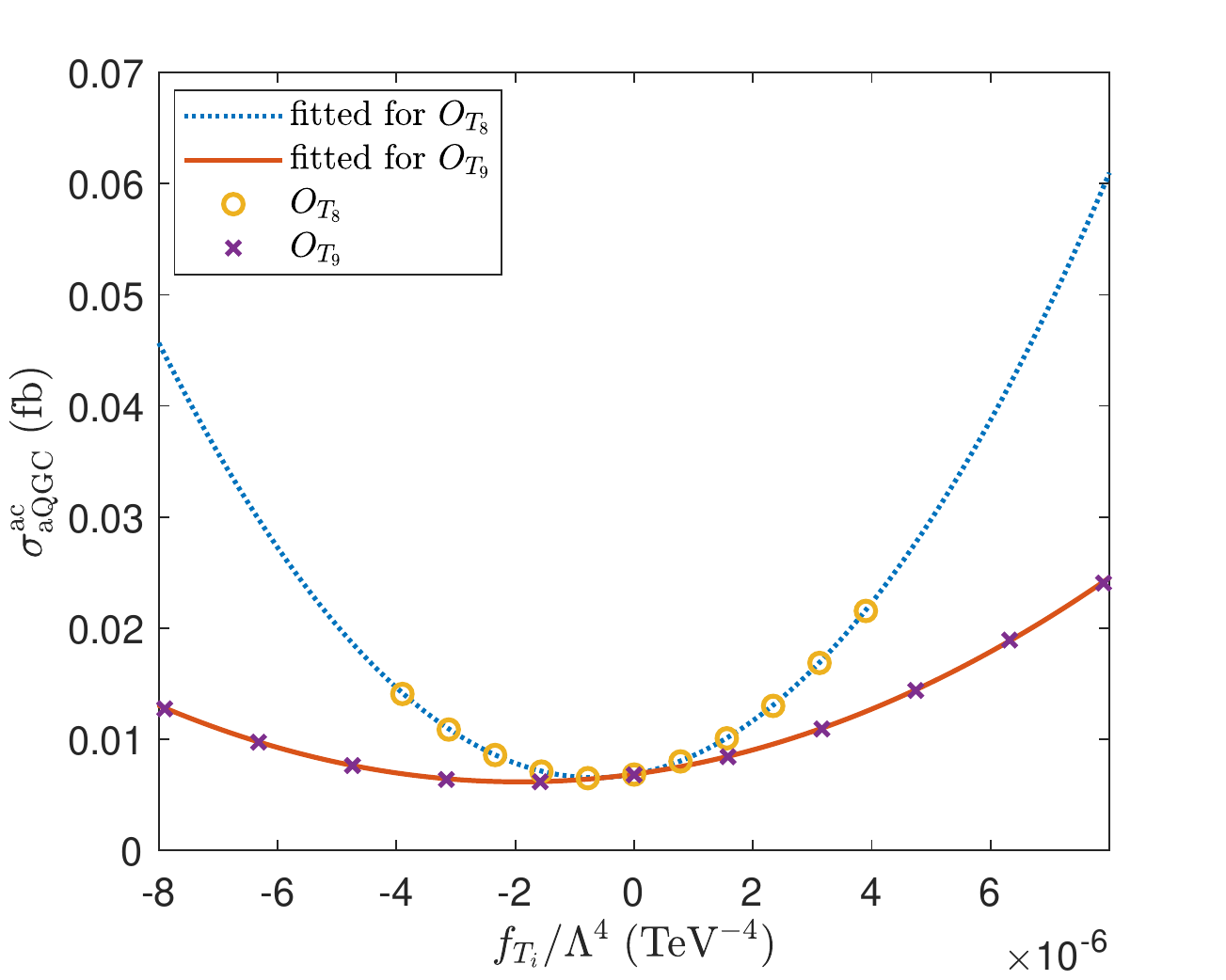}
\caption{\label{fig:css3}The calculated and fitted $\sigma_{\rm aQGC}^{\rm ac}$ as a function of $f_{T_i}/\Lambda^4$ at $\sqrt{s}=3$ TeV (row 1), 10 TeV (row 2), 14 TeV (row 3) and 30 TeV (row 4).
}
\end{center}
\end{figure*}

\begin{table}
\begin{center}
\begin{tabular}{c|c|c|c|c}
\hline
$\sqrt{s}$  & $3$ TeV & $10$ TeV & $14$ TeV & $30$ TeV \\
\hline
$O_{T_0} (O_{T_1})$ & $0.245$ & $2.722$  & $5.264$  & $24.38$ \\
\hline
$O_{T_2}$ & $0.185$ & $2.035$  & $3.977$  & $18.25$ \\
\hline
$O_{T_5} (O_{T_6})$ & $1.537$ & $17.02$  & $32.90$  & $151.70$ \\
\hline
$O_{T_7}$ & $1.140$ & $12.67$  & $24.97$  & $114.32$ \\
\hline
$O_{T_8}$ & $9.538$ & $106.37$ & $208.79$ & $956.90$ \\
\hline
$O_{T_9}$ & $7.213$ & $79.89$  & $155.97$ & $718.89$ \\
\hline
\end{tabular}
\end{center}
\caption{\label{tab.fit}The fitted values of $\hat{\sigma} _{\rm int}$~($\rm fb\times TeV^4$) for different $\sqrt{s}$, where $\hat{\sigma} _{\rm int}$ is defined in Eq.~(\ref{eq.3.3}).}
\end{table}

The total cross-sections after event selection strategy (denoted by $\sigma_{\rm aQGC}^{\rm ac}$) are obtained as a function of $f_{T_i}/\Lambda^4$ by scanning some values of coefficient with one operator at a time, and shown in Fig.~\ref{fig:css3}.
With the interference, these total cross-sections $\sigma_{\rm aQGC}^{\rm ac}$ can be fitted as bilinear functions of $f_{T_i}$, as
\begin{equation}
\begin{split}
&\sigma_{\rm aQGC}^{\rm ac} (f_{T_i})=\epsilon_{\rm SM}\sigma_{\rm SM}+ \epsilon_{O_{T_i}}\sigma _{O_{T_i}}(f_{T_i})+\frac{f_{T_i}}{\Lambda ^4} \hat{\sigma }_{\rm int} \;,
\end{split}
\label{eq.3.3}
\end{equation}
where $\epsilon_{\rm SM}$ and $\epsilon _{O_{T_i}}$ are the cut efficiencies of SM and aQGC events, respectively, and $\hat{\sigma }_{\rm int}$ is the interference parameter to be fitted. The numerical results of efficiencies are given in Table~\ref{tab.cutflow}.
The fitted results of $\hat{\sigma }_{\rm int}$ can then be obtained and are given in Table~\ref{tab.fit}.

The sensitivity of the muon colliders to the aQGC operators and the expected constraints on the coefficients can be obtained with respect to the significance defined as~\cite{Cowan:2010js,pdg}
\begin{eqnarray}
\mathcal{S}_{stat}=\sqrt{2 \left[(N_{\rm bg}+N_{s}) \ln (1+N_{s}/N_{\rm bg})-N_{s}\right]}\;,
\end{eqnarray}
where $N_s=N_{\rm aQGC}-N_{\rm SM}$ and $N_{\rm bg}=N_{\rm SM}$.
The likelihood function for $N_s$ is defined as
\begin{eqnarray}
L(N_s)={(N_{\rm bg}+N_{s})^n\over n!}e^{-(N_{\rm bg}+N_{s})}\;,
\end{eqnarray}
and one can find the maximum by setting $\partial {\rm ln}L/\partial N_s=0$. The approximate significance is then obtained by taking the square root of the likelihood ratio statistic $-2{\rm ln}(L(0)/L(N_s))$ after letting $n=N_{\rm bg}+N_{s}$~\cite{Cowan:2010js}.
The number of events $N_{\rm aQGC}$ is obtained by setting certain high-dimensional coefficient $f_{T_i}$ in SMEFT.
The number of background events $N_{\rm SM}$ denotes the SM prediction with all high-dimensional coefficients vanishing.
The integrated luminosities $\mathcal{L}$ for both the ``conservative'' and ``optimistic'' cases are considered~\cite{muoncollider5}.
As a result, we can obtain the projected sensitivity on $f_{T_i}/\Lambda^4$ by taking $2\sigma$, $3\sigma$ or $5\sigma$ significance.
The results are shown in Tables~\ref{tab.constraint1} and \ref{tab.constraint2}.
One can be seen that the tri-photon process at the muon colliders is very sensitive to $O_{T_i}$ operators, in particular the $O_{T_{8,9}}$ operators which are only relevant for the neutral aQGCs.
For the muon collider with $\sqrt{s}=3$ TeV and $\mathcal{L}=1\;{\rm ab}^{-1}$, the expected constraints at $\mathcal{S}_{stat}=2$ are about two orders of magnitude stronger than those at the $\sqrt{s}=13$ TeV LHC at $95\%$ CL listed in Table~\ref{tab.1}.
At $\sqrt{s}=30$ TeV and $\mathcal{L}=90\;{\rm ab}^{-1}$, the projected constraints would be about six orders of magnitude stronger than the $13$ TeV LHC.

Recently, a Snowmass paper investigated the searches of aQGCs through the production of $WW$ boson pairs at a muon collider with $\sqrt{s}=6$ TeV and $\mathcal{L}=4~{\rm ab}^{-1}$~\cite{Abbott:2022jqq}. They studied the $WW\nu\nu$ and $WW\mu\mu$ final states with the $W$ bosons' hadronic decay.
We also attempted to measure the aQGCs in $W^+W^-\to W^+W^-$ scattering at muon collider using artificial neural networks~\cite{Yang:2022fhw}.
In principle, the tri-photon channel can only be used to constrain $O_{T_i}$ operators, whereas $O_{M_i,S_i}$ operators can be studied in these VBS $W^+W^-$ production modes. We obtain the 95\% CL constraints on $O_{T_{0(1),2}}$ operators as $0.1447~{\rm TeV}^{-4}$ and $0.2266~{\rm TeV}^{-4}$ for $\sqrt{s}=3$ TeV and $\mathcal{L}=1~{\rm ab}^{-1}$, and $0.001219~{\rm TeV}^{-4}$ and $0.001931~{\rm TeV}^{-4}$ for $\sqrt{s}=10$ TeV and $\mathcal{L}=10~{\rm ab}^{-1}$. The upper limits from the $WW\nu\nu$ channel are $0.0030~{\rm TeV}^{-4}$ and $0.0046~{\rm TeV}^{-4}$ for $\sqrt{s}=6$ TeV and $\mathcal{L}=4~{\rm ab}^{-1}$~\cite{Abbott:2022jqq}, which are stronger than our $\sqrt{s}=3$ TeV case but less stringent than those for $\sqrt{s}=10$ TeV.
For $O_{T_{6,7}}$ operators, the expected constraints at $\sqrt{s}=3$ TeV and $\mathcal{L}=1~{\rm ab}^{-1}$ are $0.0261~{\rm TeV}^{-4}$ and $0.0398~{\rm TeV}^{-4}$ which are competitive to $0.033~{\rm TeV}^{-4}$ and $0.038~{\rm TeV}^{-4}$ from $W^+W^-\mu^+\mu^-$ channel at $\sqrt{s}=6$ TeV and $\mathcal{L}=4~{\rm ab}^{-1}$.
At $\sqrt{s}=30$ TeV, our expected constraints on $O_{T_i}$ are comparable with the results by extracting the $WWWW$ aQGC using artificial neural networks~\cite{Yang:2022fhw}.

\begin{table*}
\begin{center}
\begin{tabular}{c|c|c|c|c|c}
\hline
    & & $3$ TeV & $10$ TeV & $14$ TeV & $30$ TeV \\
    & & $1\;{\rm ab}^{-1}$ & $10\;{\rm ab}^{-1}$ & $10\;{\rm ab}^{-1}$ & $10\;{\rm ab}^{-1}$ \\
    & $\mathcal{S}_{stat}$ & $(10^{-2}~{\rm TeV^{-4}})$ & $(10^{-4}~{\rm TeV^{-4}})$ & $(10^{-4}~{\rm TeV^{-4}})$ & $(10^{-5}~{\rm TeV^{-4}})$ \\
\hline
                             & 2 & [-43.49, 14.47] & [-35.72, 12.19] & [-10.14, 4.09] & [-6.19, 3.30]  \\
$\frac{f_{T_0}(f_{T_1})}{\Lambda ^4}$
                             & 3 & [-48,57, 19.55] & [-39.98, 16.45] & [-11.50, 5.46] & [-7.21, 4.32] \\
                             & 5 & [-57.08, 28.06] & [-47.10, 23.57] & [-13.77, 7.73] & [-8.92, 6.03] \\
\hline
                             & 2 & [-108.0, 22.66] & [-87.71, 19.31] & [-24.37, 6.62]  & [-14.06, 5.65] \\
$\frac{f_{T_2}}{\Lambda ^4}$ & 3 & [-116.9, 31.59] & [-95.25, 26.85] & [-26.85,                              9.09]  & [-16.00, 7.58] \\
                             & 5 & [-132.4, 47.06] & [-108.3, 39.87] & [-31.08, 13.32] & [-19.27, 10.86] \\
\hline
                             & 2 & [-10.78, 2.61] & [-8.81, 2.22]  & [-2.45, 0.758] & [-1.44, 0.638] \\
$\frac{f_{T_5}(f_{T_6})}{\Lambda ^4}$
                             & 3 & [-11.78, 3.61] & [-9.65, 3.05]  & [-2.72, 1.03]  & [-1.65, 0.846] \\
                             & 5 & [-13.49, 5.32] & [-11.08, 4.49] & [-3.19, 1.50] & [-2.01, 1.20] \\
\hline
                             & 2 & [-27.54, 3.98] & [-22.47, 3.38] & [-6.17, 1.17] & [-3.41, 1.04] \\
$\frac{f_{T_7}}{\Lambda ^4}$ & 3 & [-29.22, 5.66] & [-23.89, 4.80] & [-6.64,                                 1.65] & [-3.79, 1.43] \\
                             & 5 & [-32.22, 8.66] & [-26.42, 7.32] & [-7.48, 2.48] & [-4.46, 2.10] \\
\hline
                             & 2 & [-1.74, 0.42] & [-1.42, 0.355] & [-0.399, 0.121] & [-0.233, 0.102] \\
$\frac{f_{T_8}}{\Lambda ^4}$ & 3 & [-1.90, 0.58] & [-1.56, 0.490] & [-0.443,                                 0.165] & [-0.267, 0.136] \\
                             & 5 & [-2.17, 0.86] & [-1.79, 0.721] & [-0.518, 0.239] & [-0.325, 0.193] \\
\hline
                             & 2 & [-4.50, 0.63] & [-3.66, 0.538] & [-1.00, 0.188] & [-0.553, 0.167] \\
$\frac{f_{T_9}}{\Lambda ^4}$ & 3 & [-4.77, 0.90] & [-3.89, 0.765] & [-1.07,                                  0.264] & [-0.615, 0.229] \\
                             & 5 & [-5.25, 1.38] & [-4.29, 1.17]  & [-1.21, 0.399] & [-0.723, 0.337] \\
\hline
\end{tabular}
\end{center}
\caption{\label{tab.constraint1}The projected sensitivities on the aQGC coefficients at the muon colliders with different c.m. energies and integrated luminosities for ``conservative'' case.
}
\end{table*}

\begin{table*}
\begin{center}
\begin{tabular}{c|c|c|c}
\hline
    & & $14$ TeV & $30$ TeV \\
    & & $20\;{\rm ab}^{-1}$ & $90\;{\rm ab}^{-1}$ \\
    & $\mathcal{S}_{stat}$  & $(10^{-5}~{\rm TeV^{-4}})$ & $(10^{-6}~{\rm TeV^{-4}})$ \\
\hline
                             & 2 & [-92.11, 31.68] & [-43.99, 15.08] \\
$\frac{f_{T_0}(f_{T_1})}{\Lambda ^4}$ & 3 & [-103.1, 42.71] & [-49.24, 20.34] \\
                             & 5 & [-121.6, 61.15] & [-58.04, 29.13] \\
\hline
                             & 2 & [-227.4, 49.89] & [-108.0, 23.88] \\
$\frac{f_{T_2}}{\Lambda ^4}$ & 3 & [-246.9, 69.37] & [-117.3, 33.19] \\
                             & 5 & [-280.6, 103.0] & [-133.4, 49.27] \\
\hline
                             & 2 & [-22.68, 5.76]  & [-10.81, 2.75] \\
$\frac{f_{T_5}(f_{T_6})}{\Lambda ^4}$
                             & 3 & [-24.86, 7.94] & [-11.85, 3.78] \\
                             & 5 & [-28.58, 11.66] & [-13.62, 5.56] \\
\hline
                             & 2 & [-58.64, 8.67] & [-27.79, 4.16] \\
$\frac{f_{T_7}}{\Lambda ^4}$ & 3 & [-62.29, 12.32] & [-29.53, 5.91] \\
                             & 5 & [-68.80, 18.83] & [-32.65, 9.02] \\
\hline
                             & 2 & [-3.70, 0.916] & [-1.76, 0.439] \\
$\frac{f_{T_8}}{\Lambda ^4}$ & 3 & [-4.05, 1.26]  & [-1.92, 0.605] \\
                             & 5 & [-4.64, 1.86]  & [-2.21, 0.890] \\
\hline
                             & 2 & [-9.49, 1.39] & [-4.52, 0.664] \\
$\frac{f_{T_9}}{\Lambda ^4}$ & 3 & [-10.08, 1.98] & [-4.80, 0.944] \\
                             & 5 & [-11.12, 3.02] & [-5.30, 1.44] \\
\hline
\end{tabular}
\end{center}
\caption{\label{tab.constraint2}The projected sensitivities on the aQGC coefficients at the muon colliders with different c.m. energies and integrated luminosities for ``optimistic'' case.
}
\end{table*}

\section{\label{sec4}Summary}

The future muon collider provides a unique chance to explore both the high-luminosity frontier and the high-energy frontier of the particle physics.
It has been realized that the muon collider is an EW gauge boson collider~\cite{muoncollider5}.
Analogous to the VBS processes at the LHC, one expects that the VBS at the muon collider is very suitable to probe the aQGCs.
Furthermore, we propose that the tri-photon process is also a very sensitive process to the $O_{T_i}$ operators contributing to aQGCs at the muon collider.

In this paper, we study the contribution of $O_{T_i}$ operators to the tri-photon process at muon colliders.
The cross-sections induced by $O_{T_i}$ operators are calculated.
We also investigate the kinematic features of the signal events from $O_{T_i}$ operators, based on which the event selection strategy is discussed. As a result, we obtain the projected sensitivities on the aQGC coefficients.
It turns out that, the tri-photon process is especially sensitive to the $O_{T_{8,9}}$ operators which only contribute to the neutral aQGCs.
At $\sqrt{s}=3$ TeV and $\mathcal{L}=1\;{\rm ab}^{-1}$, the expected constraints on the coefficients of $O_{T_{8,9}}$ can be about two orders of magnitude stronger than those at the $13$ TeV LHC.
At $\sqrt{s}=30$ TeV and $\mathcal{L}=90\;{\rm ab}^{-1}$, the projected constraints would be about six orders of magnitude stronger.

\section*{ACKNOWLEDGMENT}

\noindent
This work was supported in part by the National Natural Science Foundation of China under Grants Nos. 11905093 and 12147214, the Natural Science Foundation of the Liaoning Scientific Committee No.~LJKZ0978 and the Outstanding Research Cultivation Program of Liaoning Normal University (No.21GDL004).
T.L. is supported by the National Natural Science Foundation of China (Grants No. 11975129, 12035008) and ``the Fundamental Research Funds for the Central Universities'', Nankai University (Grant No. 63196013).

\bibliography{triphoton}
\bibliographystyle{JHEP}

\end{document}